\documentclass{appolb}
\usepackage{graphicx}
\usepackage{amssymb}
   \usepackage{amsmath}
    \usepackage{amsfonts}
     \usepackage{epsfig}
      \usepackage{bm}
\def\Pom{{\bf I\!P}}
\newcommand{\ket}[1]{| {#1} \rangle}
\newcommand{\bra}[1]{\langle {#1} |}

\newcommand{\bq}{{\bf{q}}}
\newcommand{\bp}{{\bf{p}}}

\begin{document}
\title{Some new aspects of quarkonia production at the LHC
\thanks{Presented at the Epiphany 2017 conference}
}

%
\author{Antoni Szczurek
\address{Institute of Nuclear Physics Polish Academy of Sciences, PL-31342 Krak\'ow}
\\
{Anna Cisek}
\address{
Faculty of Mathematics and Natural Sciences, University of Rzeszow, ul.
Pigonia 1, 35-310 Rzesz\'ow}
\\
{Wolfgang Sch\"afer}
\address{Institute of Nuclear Physics Polish Academy of Sciences, PL-31342 Krak\'ow}
}

\maketitle
\begin{abstract}
We discuss some different new aspects of $J/\psi$ meson production
in exclusive, semi-exclusive and inclusive processes.
We finish with a short discussion of double $J/\psi$ production.
We point out some new results obtained recently by our group
and discuss some open issues.
\end{abstract}
  
\section{Introduction}

The $J/\psi$ meson is interesting due to its simple structure 
($c \bar c$ state). However, the mechanism of its production is not
always well understood. There are different subfields of high energy
physics that deal with the production of $J/\psi$ meson. Our group
participated in different fields and it is the aim of this talk 
(presentation) to show some progress in the different fields.

The first topic is connected with the exclusive production of 
$J/\psi$ mesons in the $p p \to p p J/\psi$ reaction. 
There the main production mechanism is a photon-pomeron fusion. 
The same type of process was studied some years ago in the context 
of $\gamma p \to J/\psi p$ reactions measured at HERA \cite{INS}. 
The pomeron is a key word which is understood and used 
by different authors often very differently. 
At high energies, which corresponds to small fractions
of gluon longitudinal momentum fractions, a problem of gluon saturation
may be important.
The problem with the current experiment is that so far
it was not possible to control the exclusivity of the process.
Therefore recently we have performed more detailed studies
of semiexclusive processes that contribute in experimental environment
but strictly speaking are not exclusive in a rigorous sense of the word.

The inclusive production of $J/\psi$ mesons is a long standing problem
in spite that there are well formulated nonrelativistic pQCD rules.
Forward production of $J/\psi$ mesons may be also related to the
phenomenon of gluon saturation.
Recent years the luminosity of the running experiments was increased
and allowed to measure even two $J/\psi$ mesons in one event. 
Here a new mechanism, called double parton (or multiple parton)
scattering, may appear at high energies, where the density of partons,
in particular of gluon becomes large. In contrast to single parton
scattering the formalism of double (or multiple) scattering is
not under similar theoretical precision.
While there was a relatively good agreement for experiments where
low transverse momenta of $J/\psi$ mesons were involved (e.g. LHCb),
the cases where cuts on transverse momenta are (were) more problematic
and large disagreement of theoretical calculation with the data
was observed unless the constant known as effective cross section
was considerably lowered. Below we wish to discuss some mechanisms
that were neglected and may be important in good understanding
of the situation.

\section{Exclusive $p p \to p p J/\psi$ reaction}

\subsection{Sketch of the theoretical methods}

\begin{figure}
\begin{center}
\includegraphics[height=5.0cm]{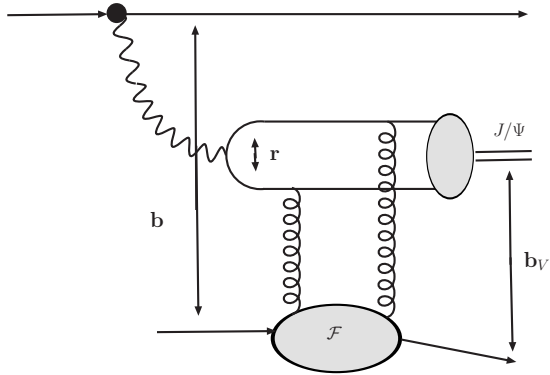}
\end{center}
\caption {A sketch of dynamics of the $\gamma p \to J/\psi p$ process.}
\label{fig:gammap_Vp}
\end{figure}

The imaginary part of the forward amplitude sketched in 
Fig.\ref{fig:gammap_Vp} 
can be written as \cite{INS}:

\begin{eqnarray}
\Im m \, {\cal M}_{T}(W,\Delta^2 = 0,Q^{2}=0) =
W^2 \frac{c_v \sqrt{4 \pi \alpha_{em}}}{4 \pi^2} \, 2 \, 
 \int_0^1 \frac{dz}{z(1-z)}
\int_0^\infty \pi dk^2 \psi_V(z,k^2)
\nonumber \\
\int_0^\infty
 {\pi d\kappa^2 \over \kappa^4} \alpha_S(q^2) {\cal{F}}(x_{\rm eff},\kappa^2)
\Big( A_0(z,k^2) \; W_0(k^2,\kappa^2) 
     + A_1(z,k^2) \; W_1(k^2,\kappa^2)
\Big) \, ,
\end{eqnarray}

The full amplitude, at finite momentum transfer is given by:
\begin{eqnarray}
{\cal M}(W,\Delta^2) = (i + \rho) \, \Im m {\cal M}(W,\Delta^2=0,Q^{2}=0)
\, \exp(-B(W) \Delta^2/2) \, ,
\label{full_amp}
\end{eqnarray}
where the real part of the amplitude is restored from analyticity,
\begin{eqnarray}
\rho = {\Re e {\cal M} \over \Im m {\cal M}} =  
\tan \Big ( {\pi \over 2} \, { \partial \log \Big( \Im m {\cal M}/W^2 \Big) \over \partial \log W^2 } \Big) \, .
\end{eqnarray}
Above $B(W)$ is a slope parameter which in general depends on the photon-proton
center-of-mass energy and is parametrized in the present analysis as:
\begin{eqnarray}
B(W) = b_0 + 2 \alpha'_{eff} \log \Big( {W^2 \over W^2_0} \Big) \, .
\end{eqnarray}

\begin{figure}
\begin{center}
\includegraphics[height=4.5cm]{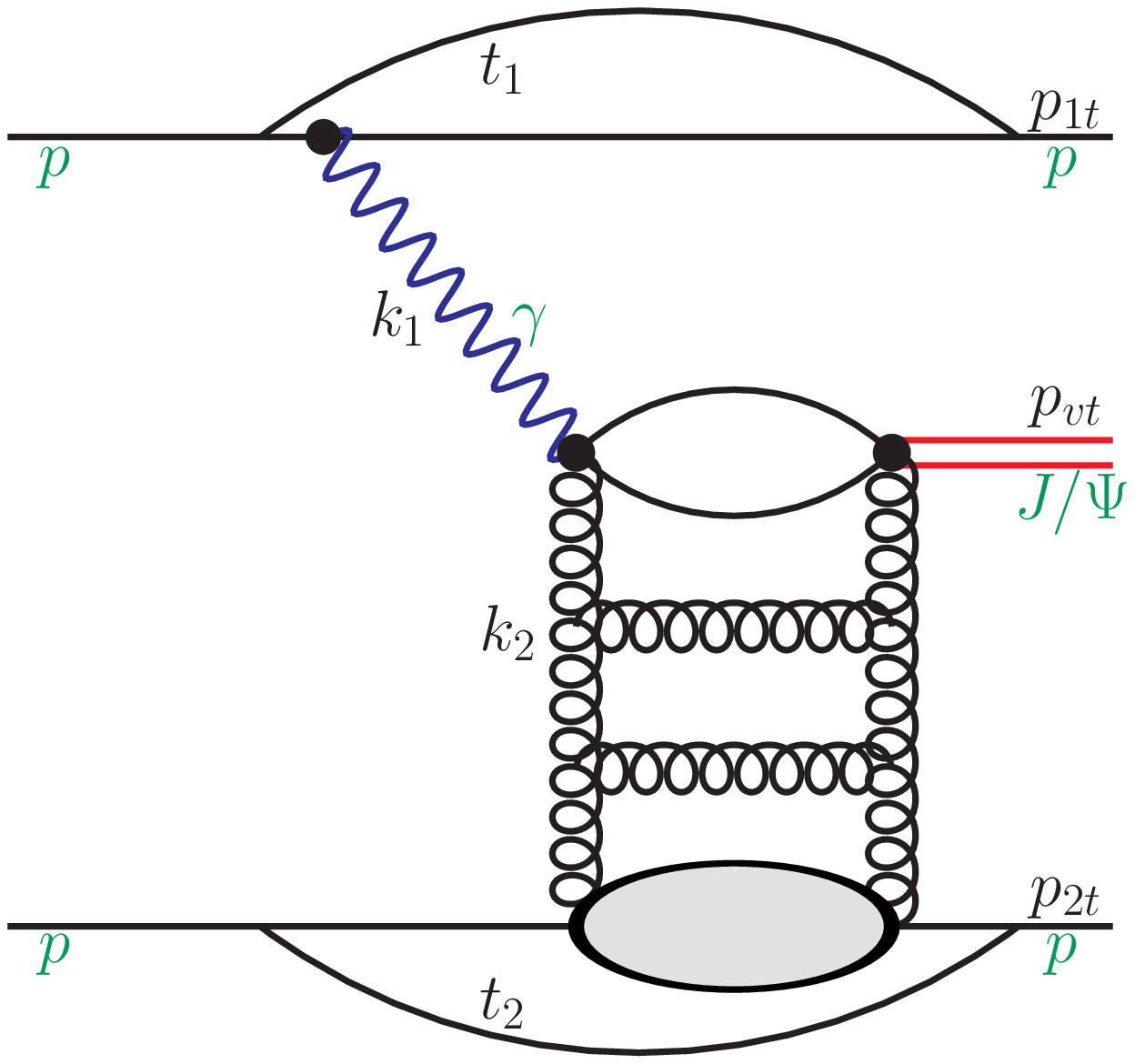}
\includegraphics[height=4.5cm]{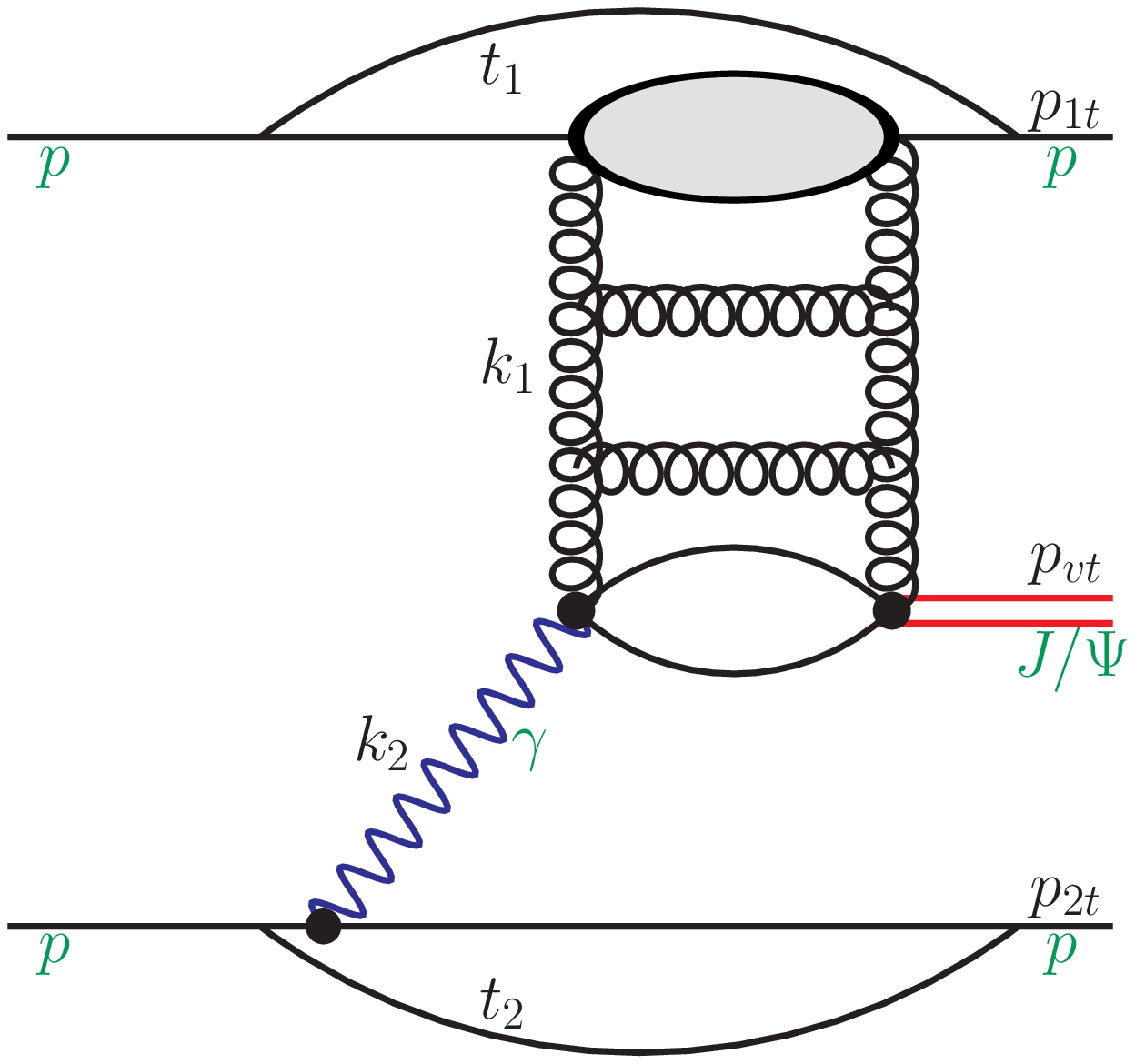}
\end{center}
\caption{Two mechanisms of exclusive $J/\psi$ meson production
at the Born level.}
\label{fig:pp_ppjpsi_Born}
\end{figure}

The full Born amplitude for the $ pp \to pVp$ process 
(see Fig.\ref{fig:pp_ppjpsi_Born}) can be written as:
\begin{eqnarray}
{\cal M}_{h_1 h_2 \to h_1 h_2 V}^
{\lambda_1 \lambda_2 \to \lambda'_1 \lambda'_2 \lambda_V}(s,s_1,s_2,t_1,t_2) =
{\cal M}_{\gamma \Pom} + {\cal M}_{\Pom \gamma} \nonumber \\
= \bra{p_1', \lambda_1'} J_\mu \ket{p_1, \lambda_1} 
\epsilon_{\mu}^*(q_1,\lambda_V) {\sqrt{ 4 \pi \alpha_{em}} \over t_1}
{\cal M}_{\gamma^* h_2 \to V h_2}^{\lambda_{\gamma^*} \lambda_2 \to \lambda_V \lambda_2}
(s_2,t_2,Q_1^2)   \nonumber \\
 + \bra{p_2', \lambda_2'} J_\mu \ket{p_2, \lambda_2} 
\epsilon_{\mu}^*(q_2,\lambda_V)  {\sqrt{ 4 \pi \alpha_{em}} \over t_2}
{\cal M}_{\gamma^* h_1 \to V h_1}^{\lambda_{\gamma^*} \lambda_1 \to \lambda_V \lambda_1}
(s_1,t_1,Q_2^2)  \, .
\label{Two_to_Three}
\end{eqnarray}
The electromagnetic transition matrix elements contain both
helicity conserving and helicity flip components \cite{CSS2015}. 

\begin{figure}
\begin{center}
\includegraphics[height=4.5cm]{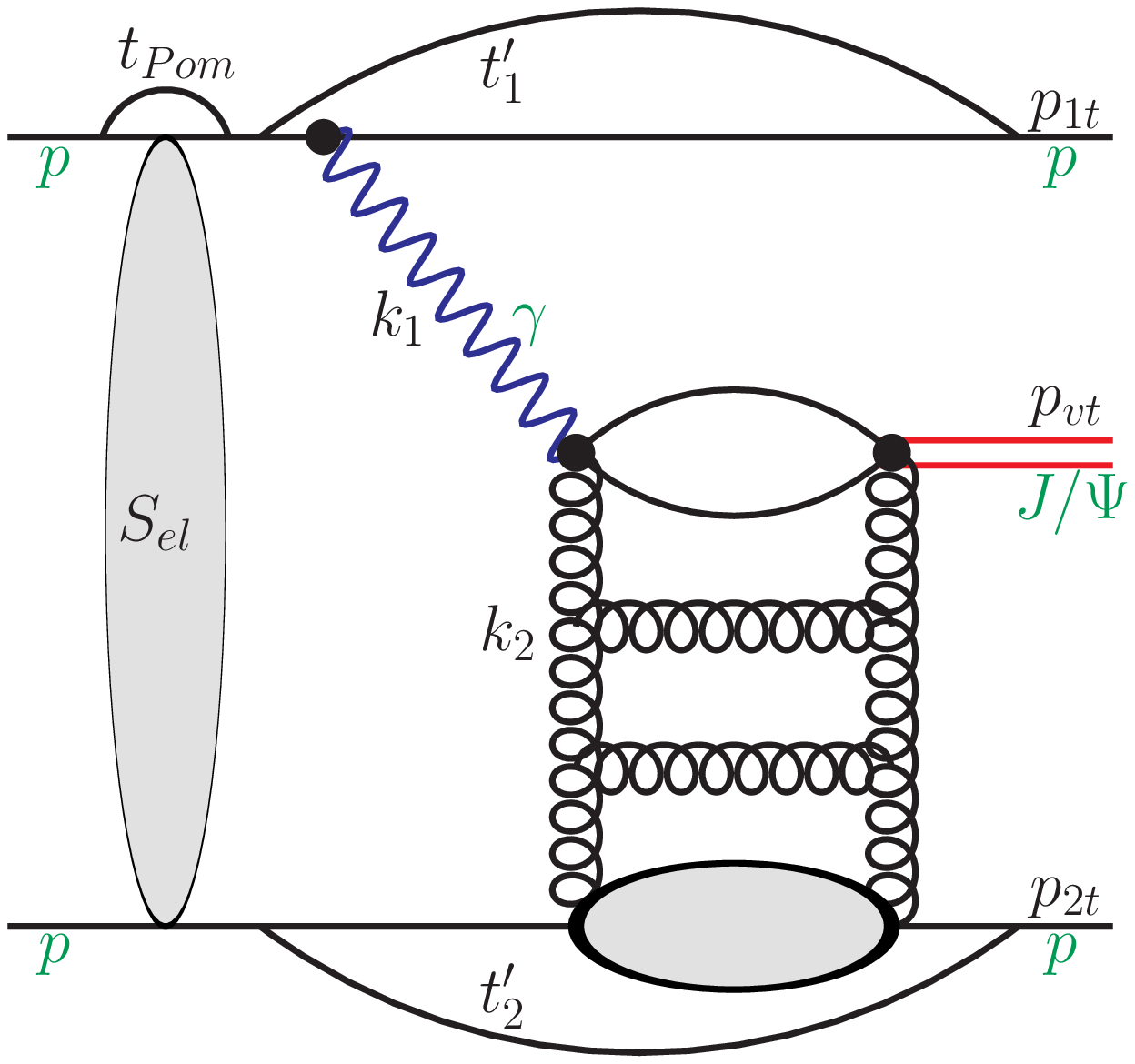}
\includegraphics[height=4.5cm]{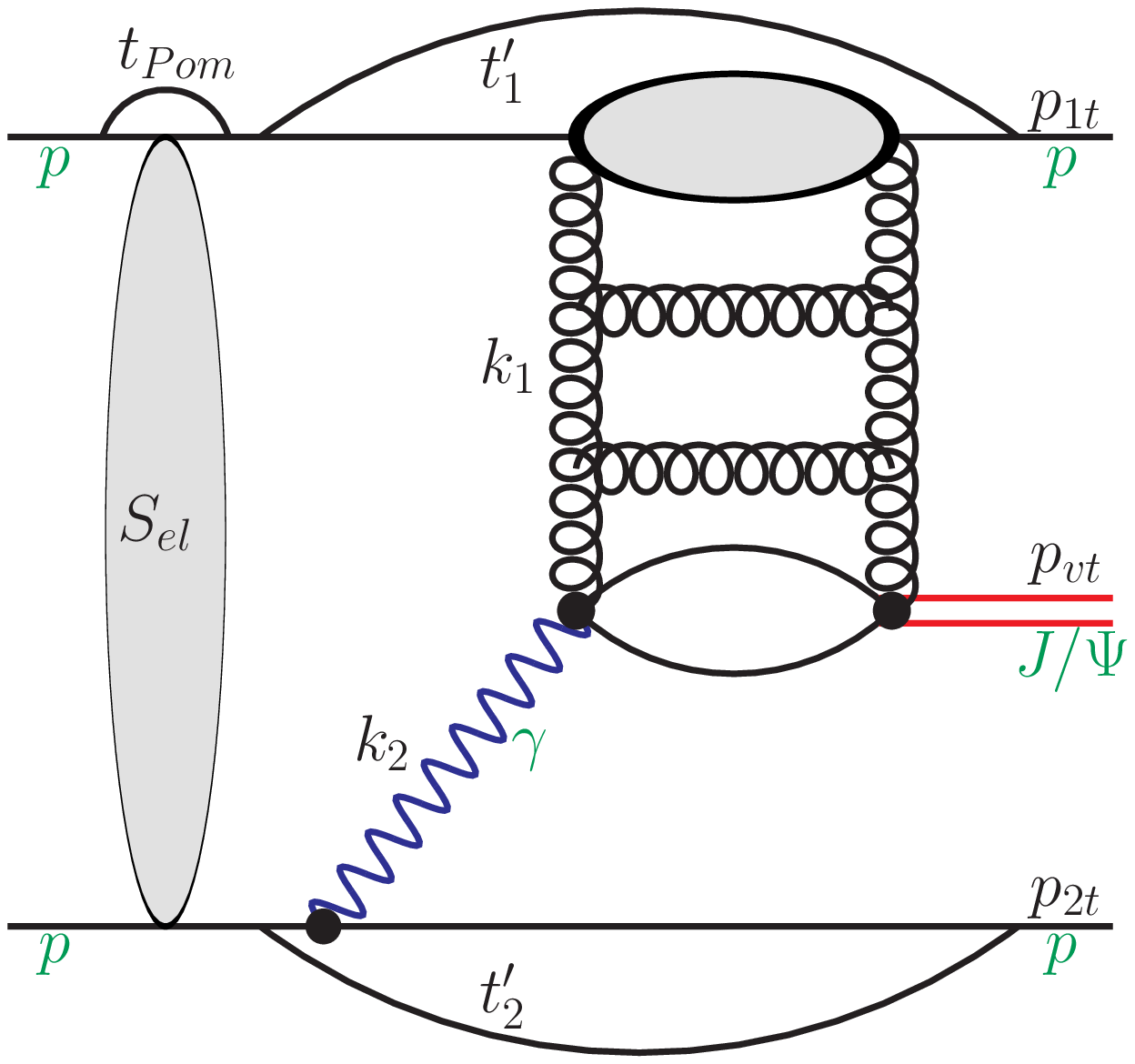}
\end{center}
\caption{A sketch of the absorption effects
that lead to a decrease of the Born cross sections.}
\label{fig:pp_ppjpsi_absorption}
\end{figure}

The effects of absorption are illustrated in 
Fig.\ref{fig:pp_ppjpsi_absorption}. The relevant formalism is
described e.g. in Ref.\cite{SS2007}.

\subsection{Selected results}

\begin{figure}
\begin{center}
\includegraphics[width=4cm]{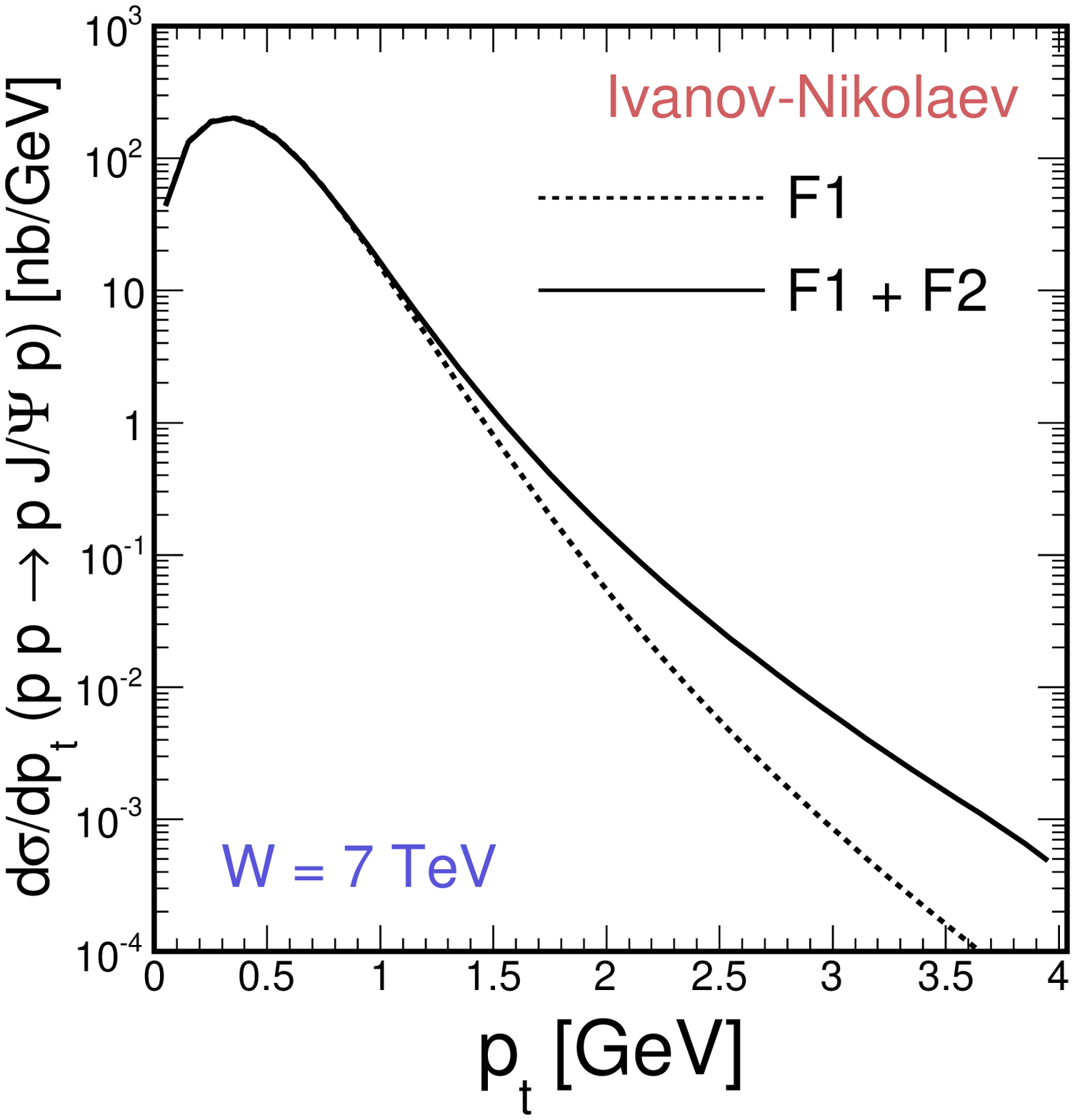}
\includegraphics[width=4cm]{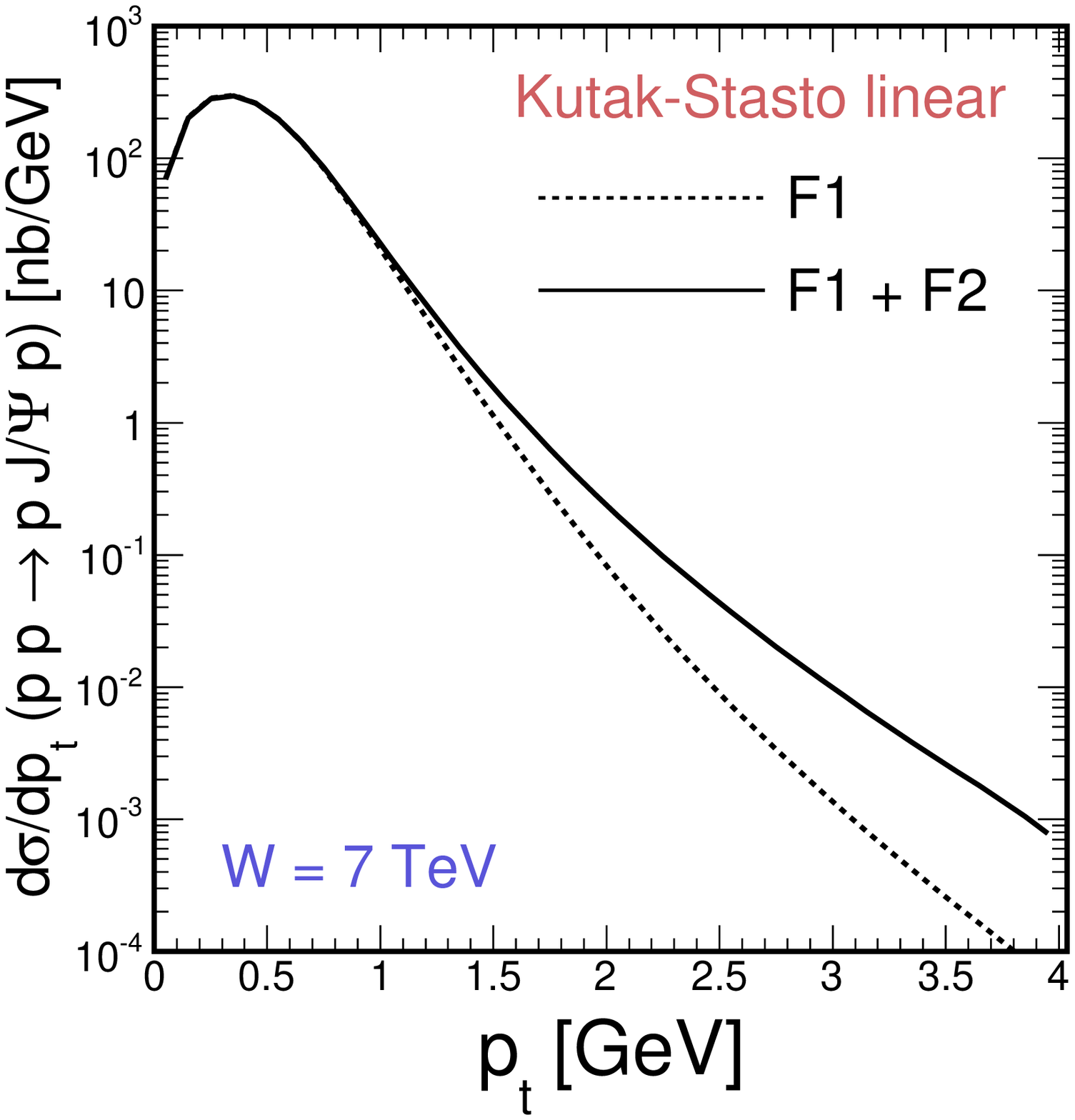}
\includegraphics[width=4cm]{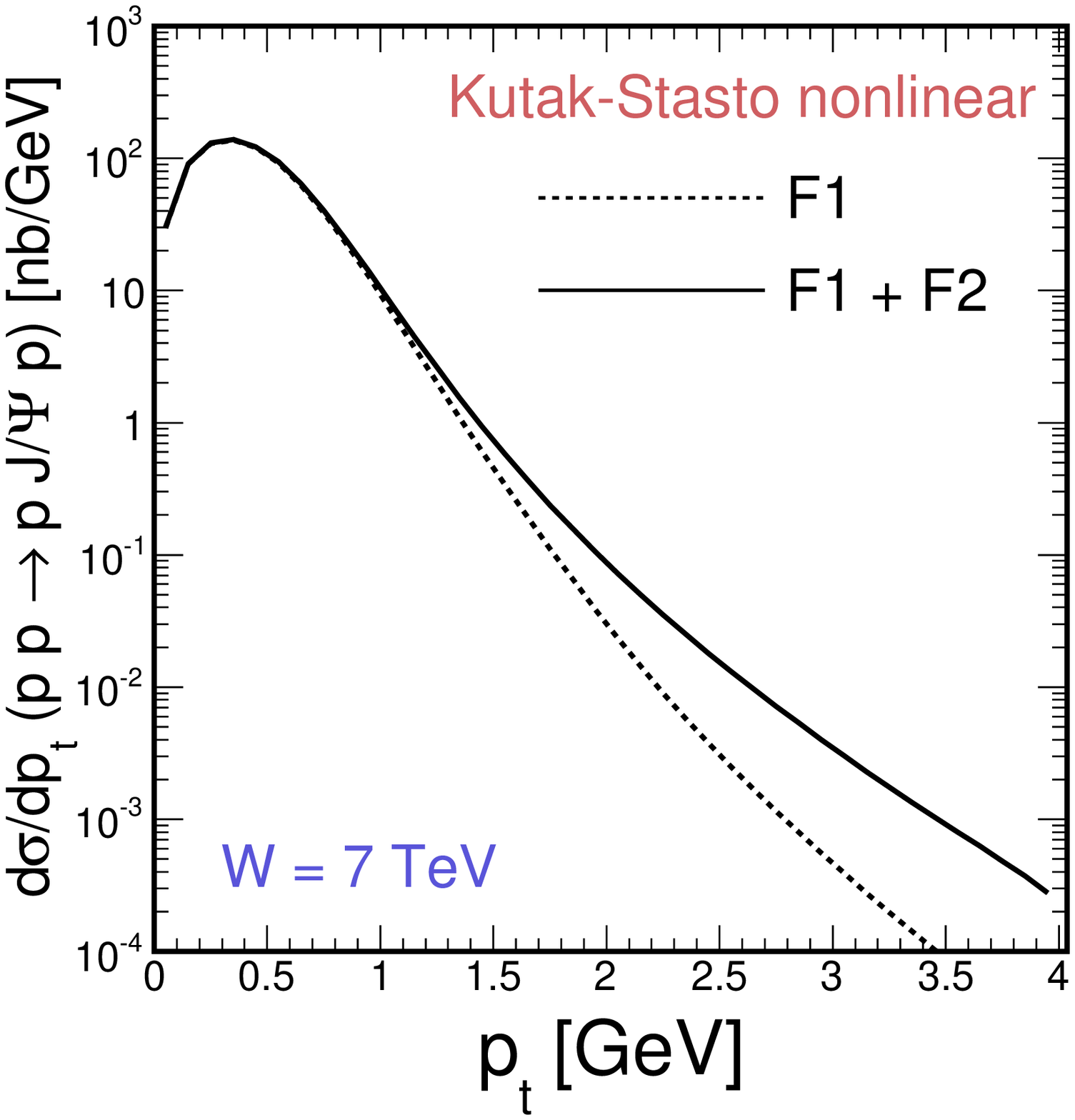}
\end{center}
\caption
{Transverse momentum distributions of exclusively produced
$J/\psi$ mesons for three different unintegrated gluon distributions.
We compare results with Dirac and Dirac+Pauli electromagnetic form factor.
}
\label{fig:dsig_dpt_jpsi_Dirac_vs_Pauli}
\end{figure}

Most of the calculations in the literature neglect the anomalous
photon coupling to proton quantified by the Pauli coupling.
In Fig.\ref{fig:dsig_dpt_jpsi_Dirac_vs_Pauli} we illustrate the role of
the Pauli form factor on the transverse momentum distribution of
$J/\psi$ mesons. Inclusion of Pauli form factor considerably enhances 
the cross section at large transverse momenta of $J/\psi$ meson.

\begin{figure}
\begin{center}
\includegraphics[width=4.1cm]{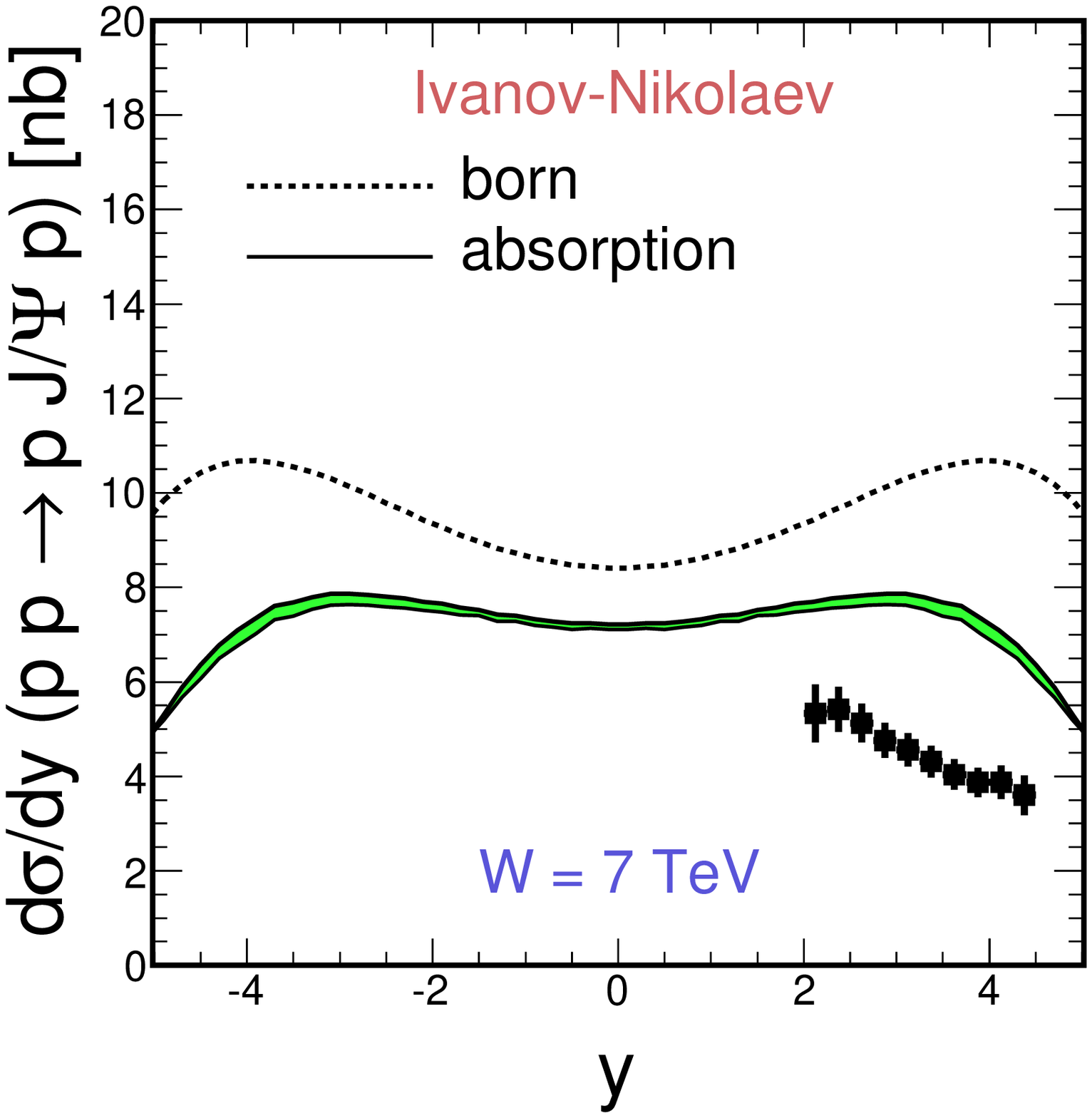}
\includegraphics[width=4.1cm]{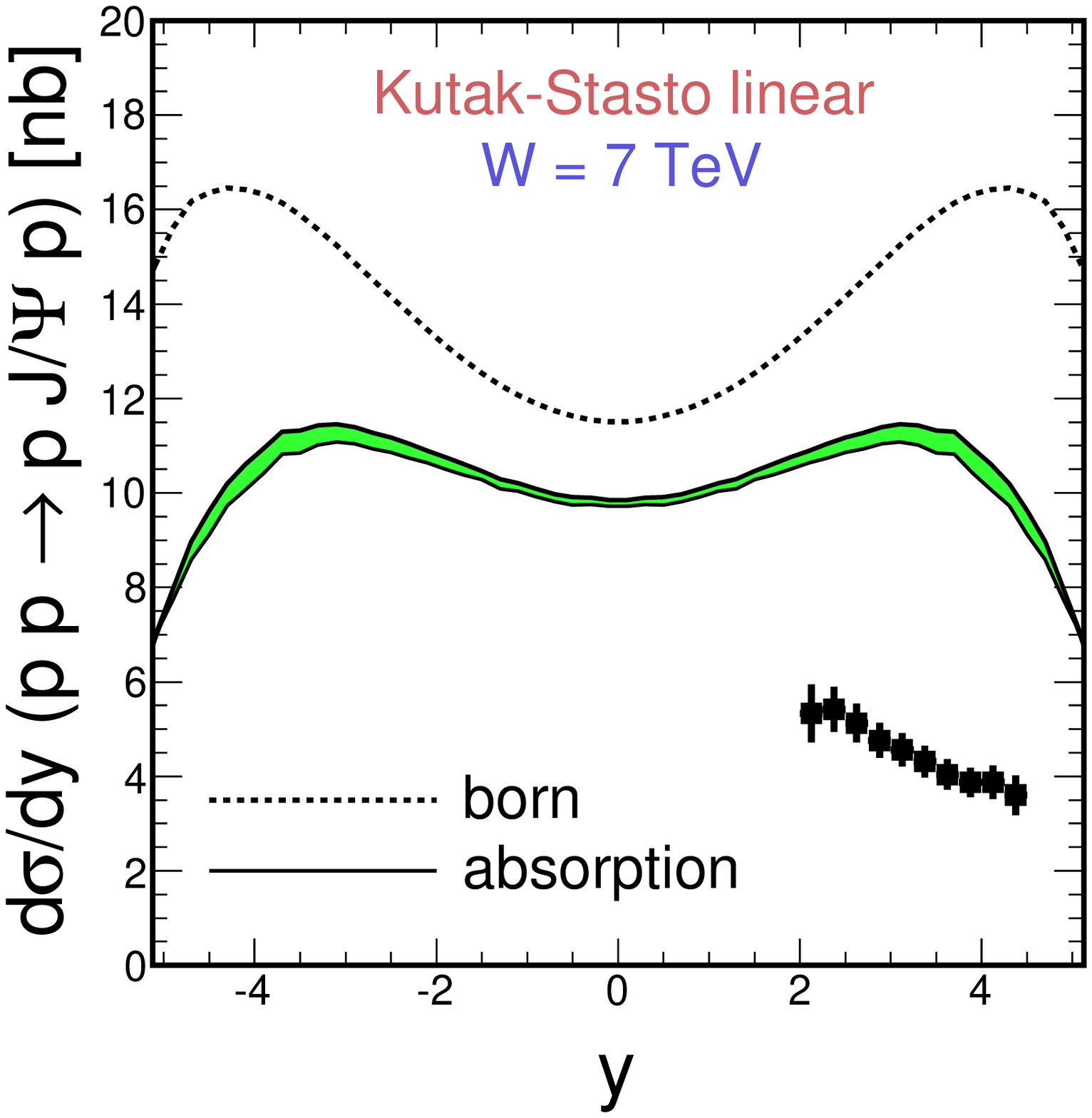}
\includegraphics[width=4.1cm]{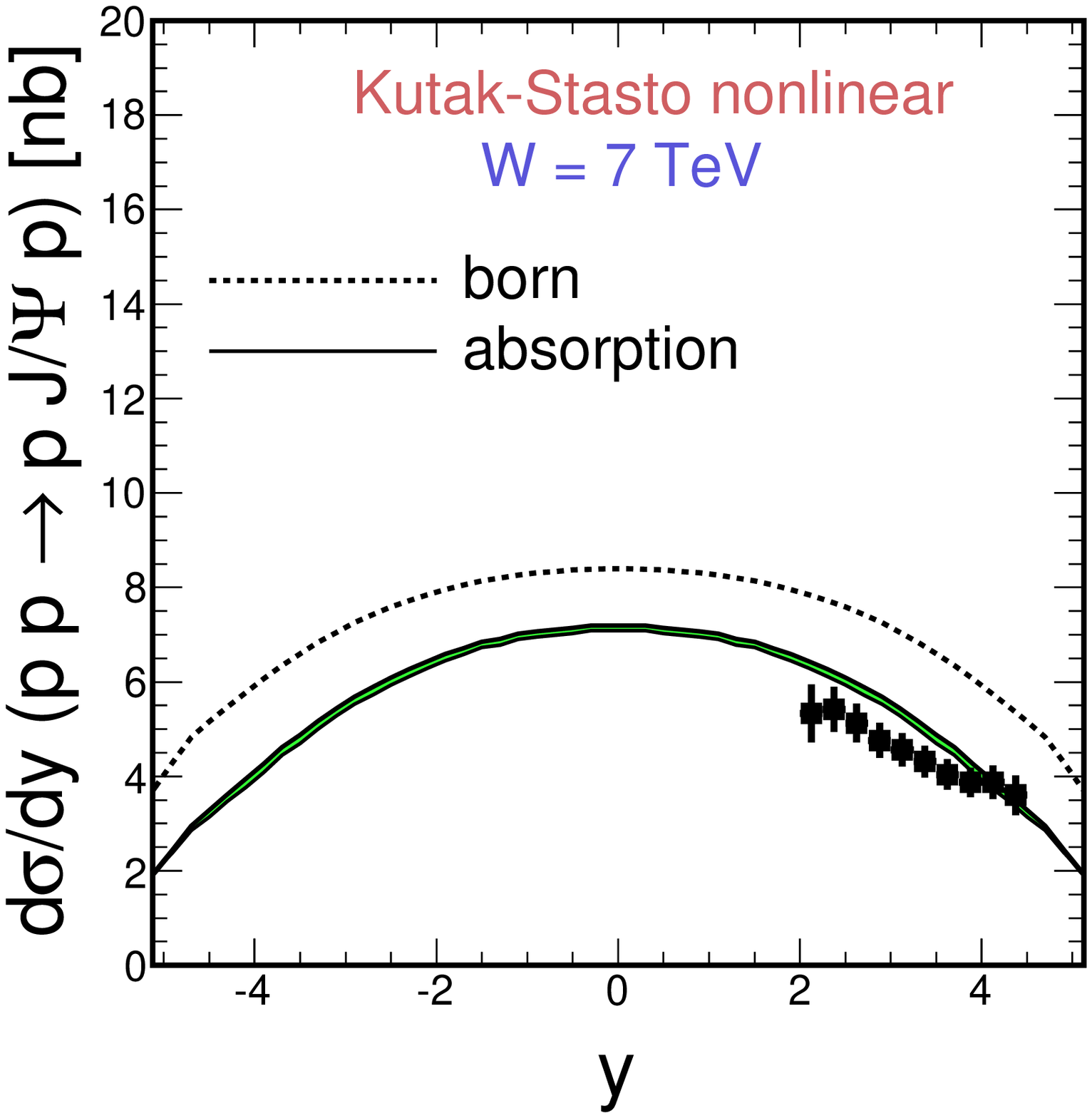}
\end{center}
\caption{Rapidity distributions of exclusively produced $J/\psi$ mesons
for three different unintegrated gluon distributions at $\sqrt{s}$ = 7
TeV.}
\label{fig:dsig_dy_jpsi}
\end{figure}

In Fig.\ref{fig:dsig_dy_jpsi} we show rapidity distributions for
different unintegrated gluon distribution function (UGDF). 
The results strongly depend on UGDF used.
The best result is achieved with UGDFs that include nonlinear effects
(here a nonlinear Kutak-Stasto UGDF was used). This may be connected
with the onset of gluon saturation, a phenomenon expected to show up 
in processes with small gluon longitudinal momentum fraction
but never unambiguously identified.

\begin{figure}
\begin{center}
\includegraphics[width=4.1cm]{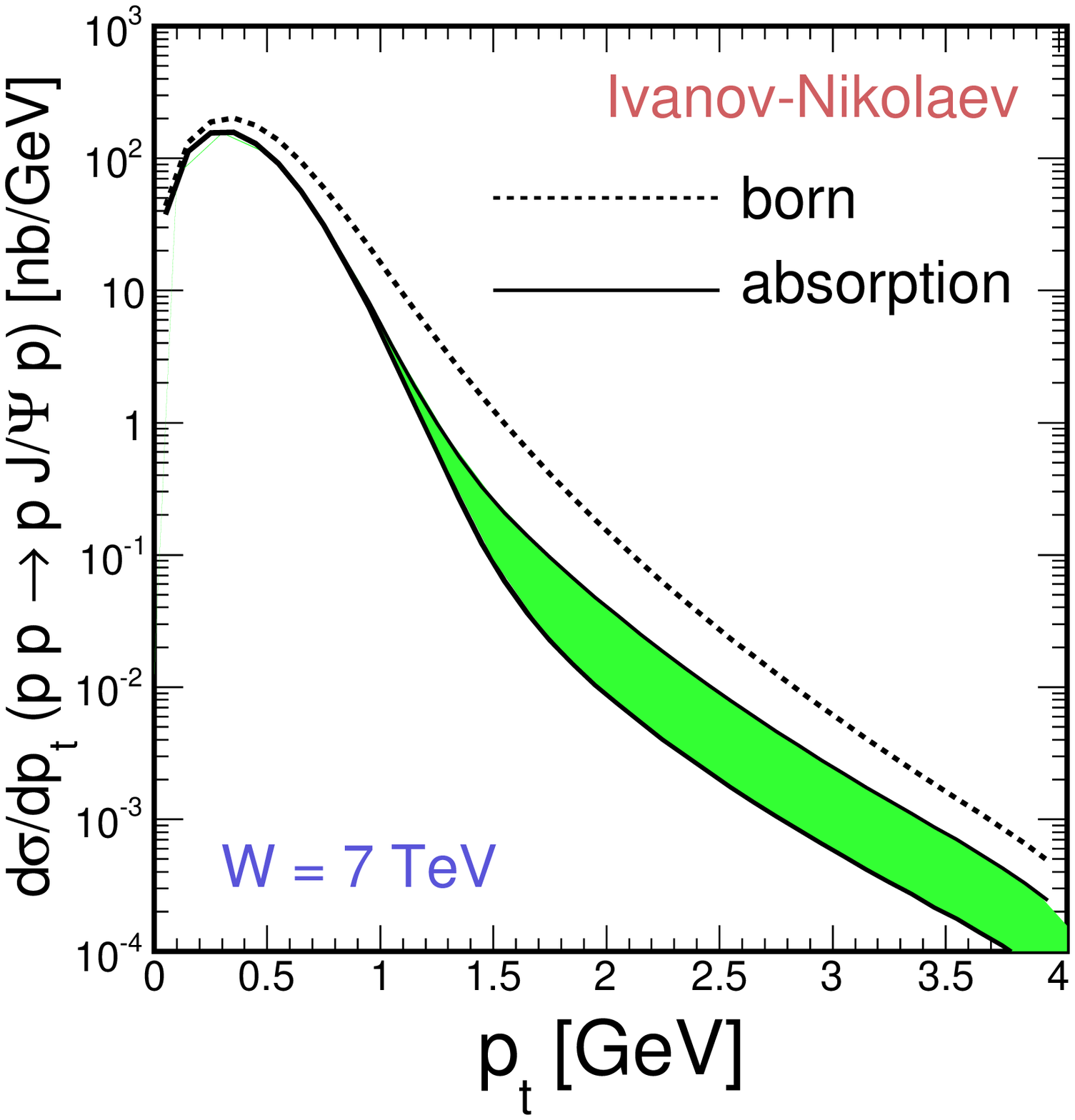}
\includegraphics[width=4.1cm]{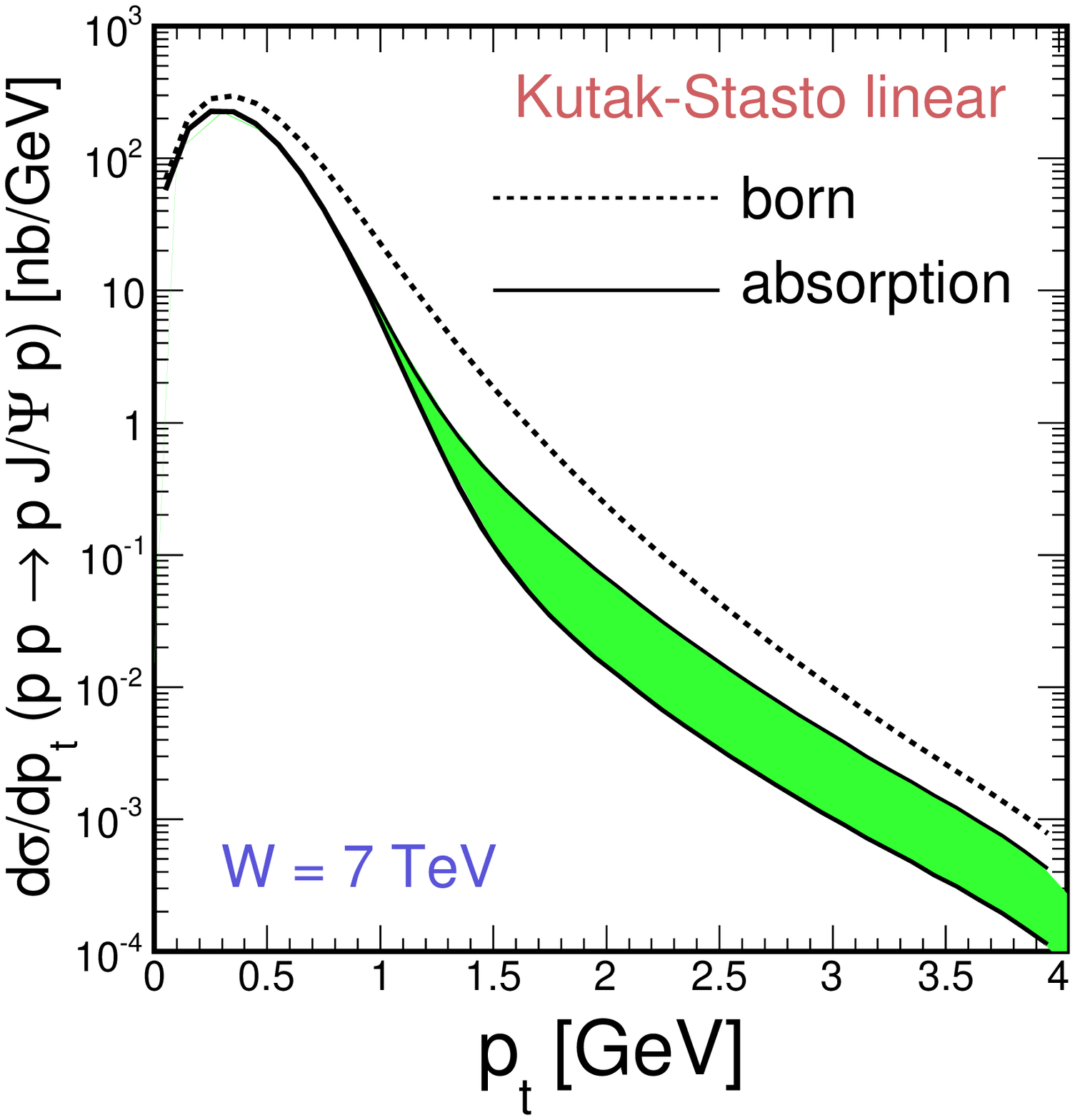}
\includegraphics[width=4.1cm]{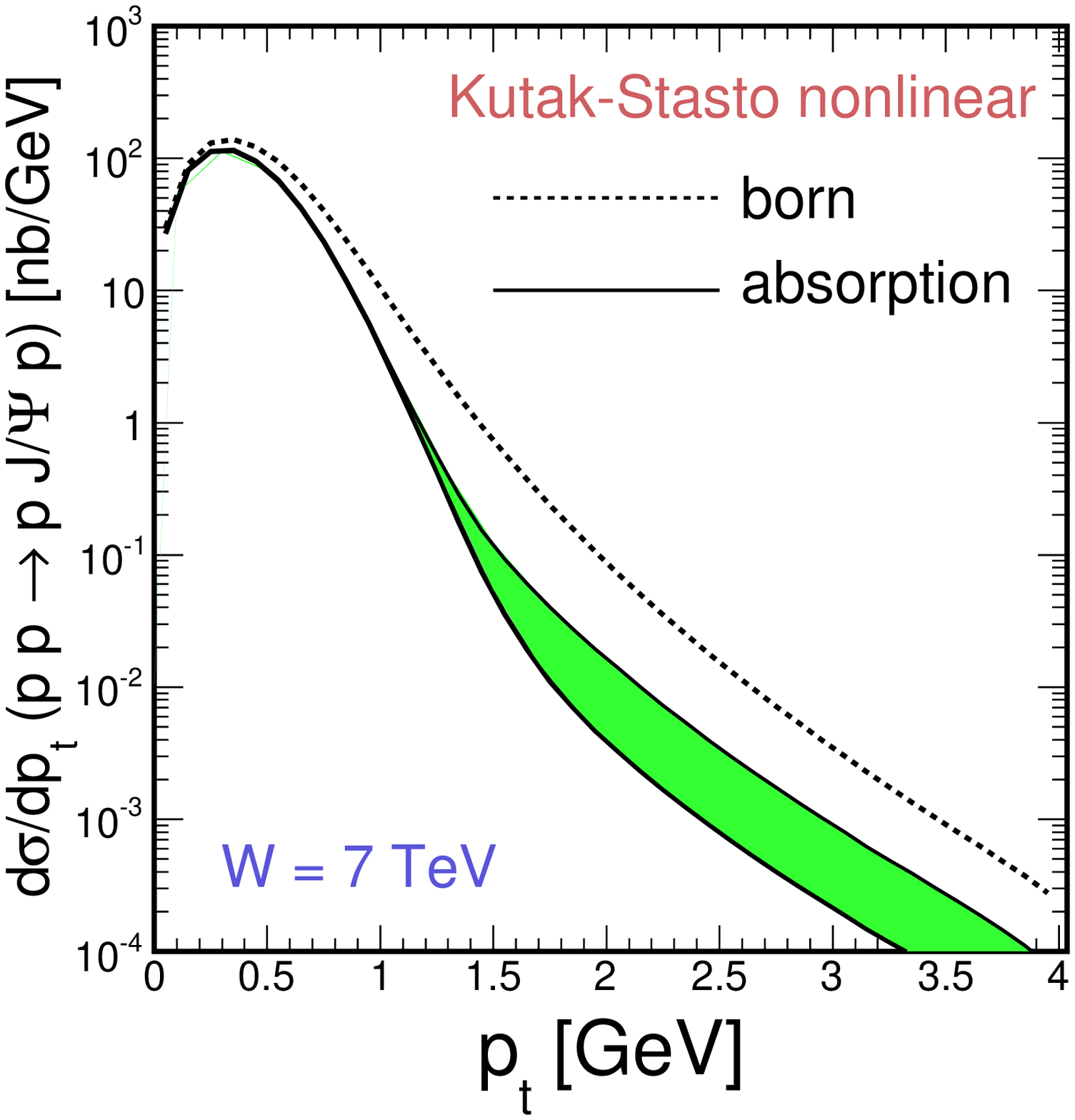}
\end{center}
\caption{Transverse momentum distributions of exclusively produced
$J/\psi$ mesons for three different unintegrated gluon distributions.}
The shaded (green online) band represents typical uncertainties 
in calculating absorption effects as described in the text.
\label{fig:dsig_dpt_jpsi_absorption}
\end{figure}

The role of absorption effects is illustrated in 
Fig.\ref{fig:dsig_dpt_jpsi_absorption}. The absorption effects
modify the shape of transverse momentum distributions.
Therefore a careful treatment of absorption effects is very
important in the context of possible searches for odderon exchange
discussed in the literature.

In Ref.\cite{CSS2015} we have shown similar effects for $\psi'$ meson
production. Here we show only the ratios of the cross section
for $\psi'$ to the cross section for $J/\psi$ meson production
as a function of $J/\psi$ ($\psi'$) rapidity.
A good agreement is achieved for the Gaussian wave functions.
In contrast, approaches that do not use wave functions
cannot give a reasonable description of the data.

\begin{figure}
\begin{center}
\includegraphics[width=4.1cm]{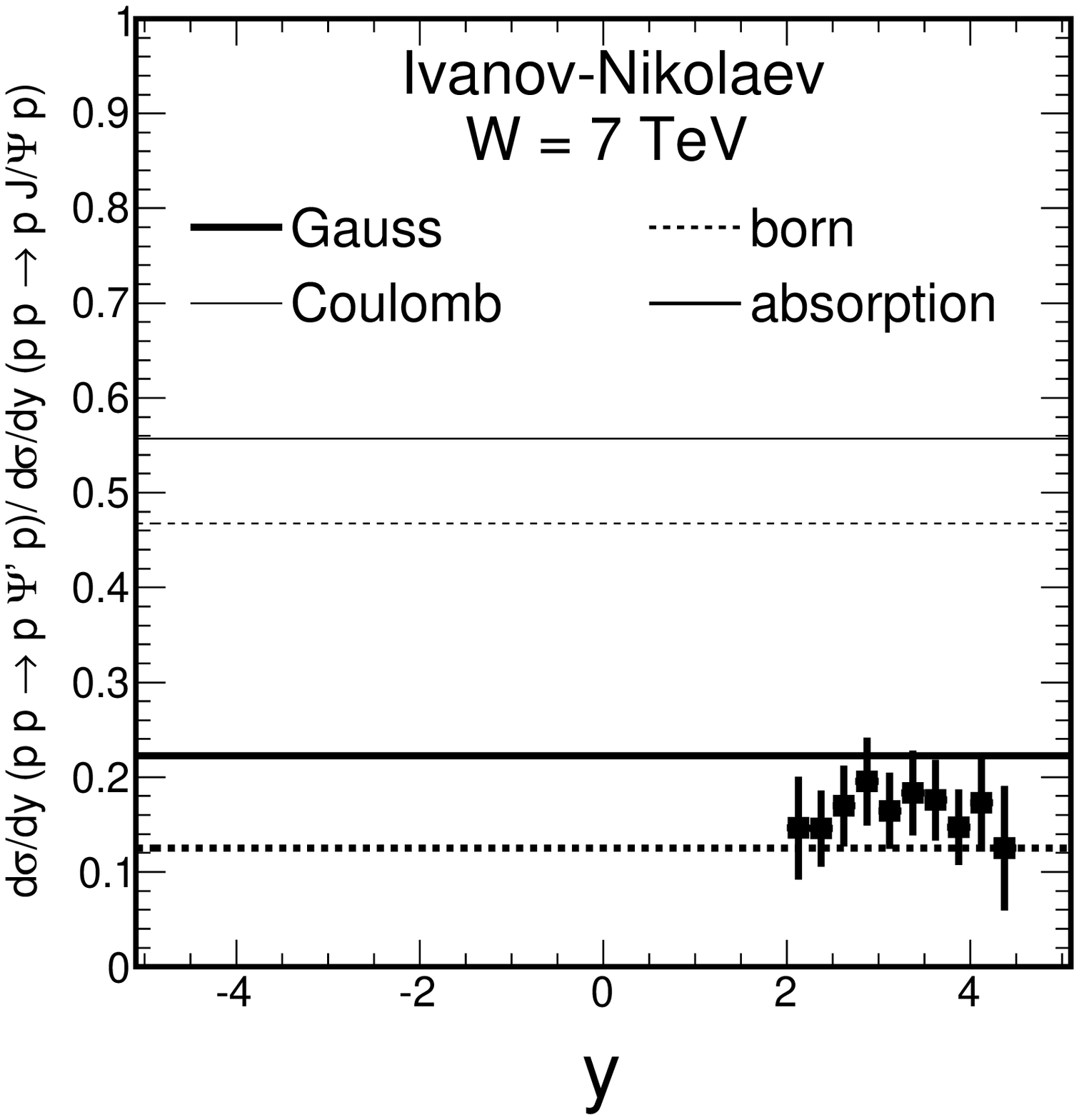}
\includegraphics[width=4.1cm]{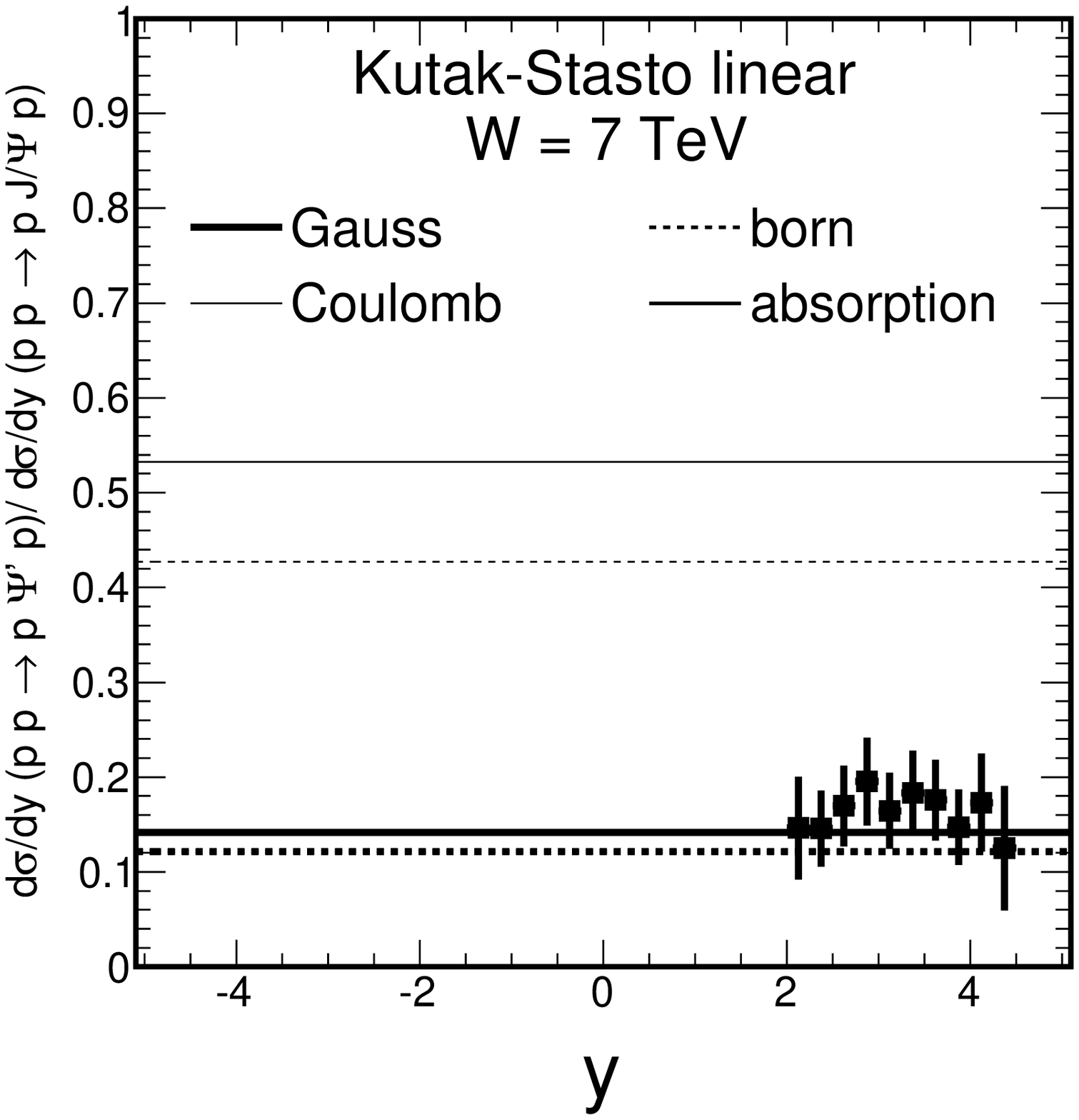}
\includegraphics[width=4.1cm]{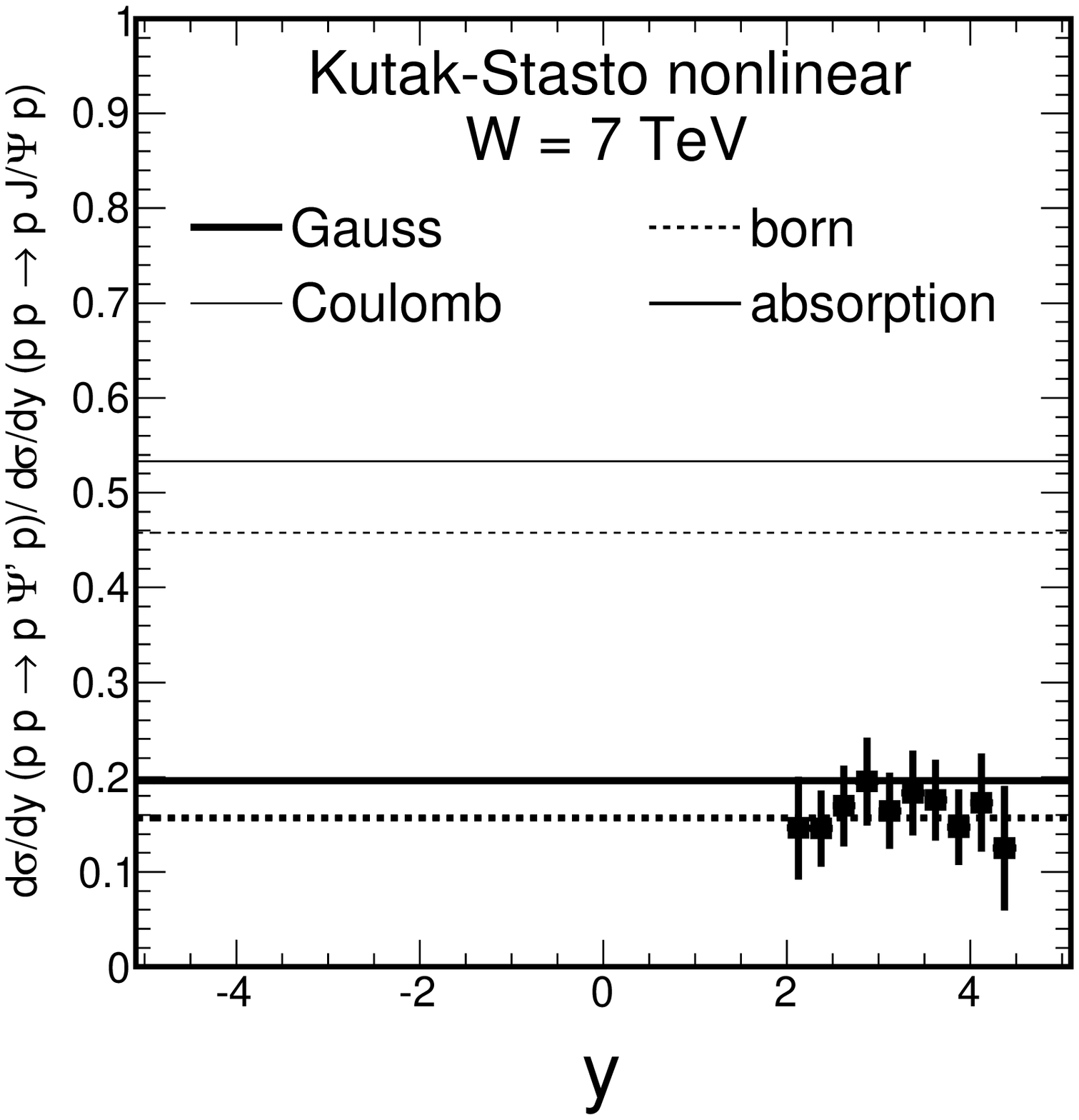}
\end{center}
\caption{The ratio of the cross section for $\psi'$ to that for
the $J/\psi$ meson.}
\end{figure}

\section{Semiexclusive production of $J/\psi$ mesons
with proton dissociation}

\subsection{Sketch of the theoretical methods}

\begin{figure}
\begin{center}
\includegraphics[width=4.5cm]{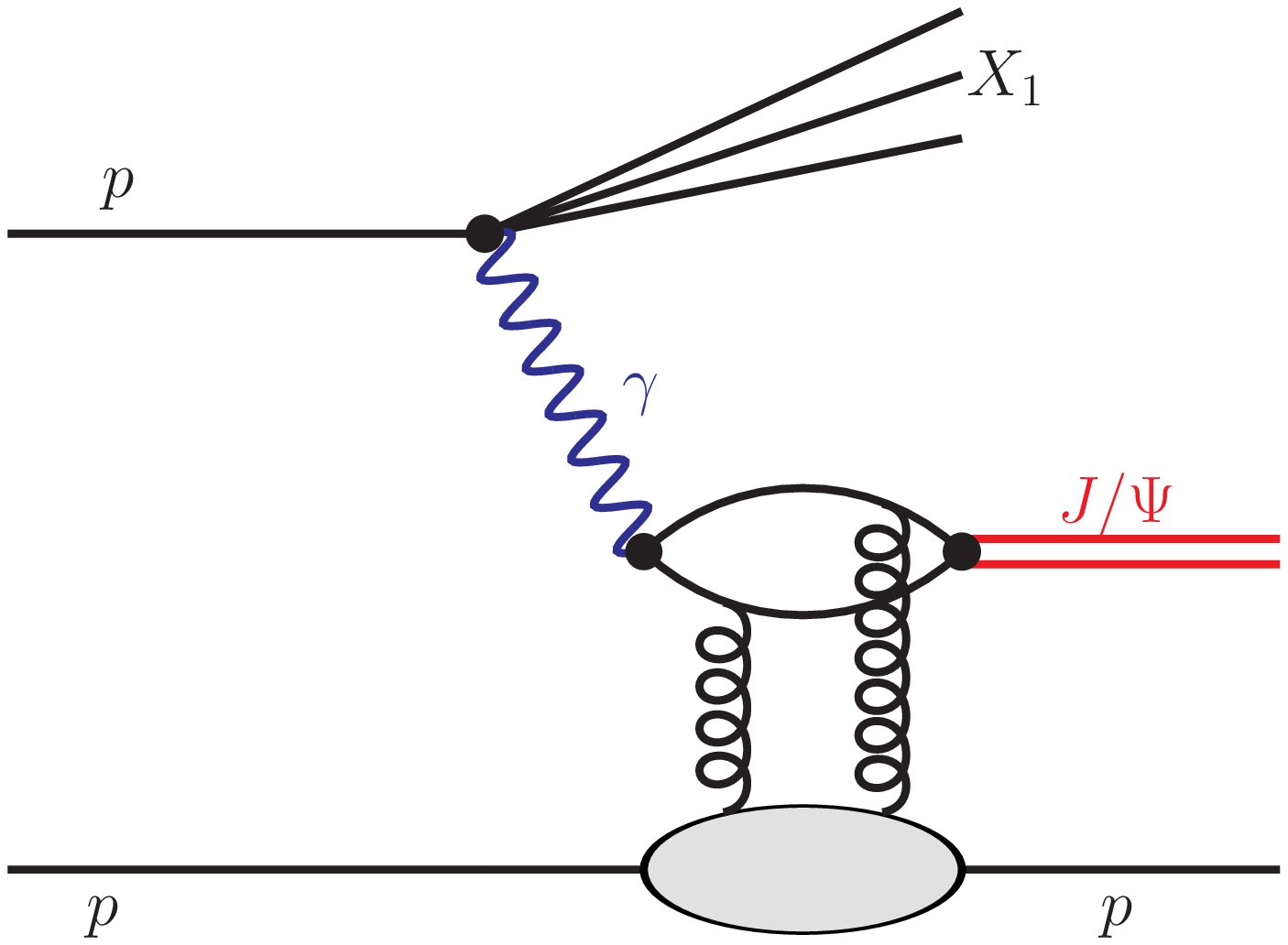}
\includegraphics[width=4.5cm]{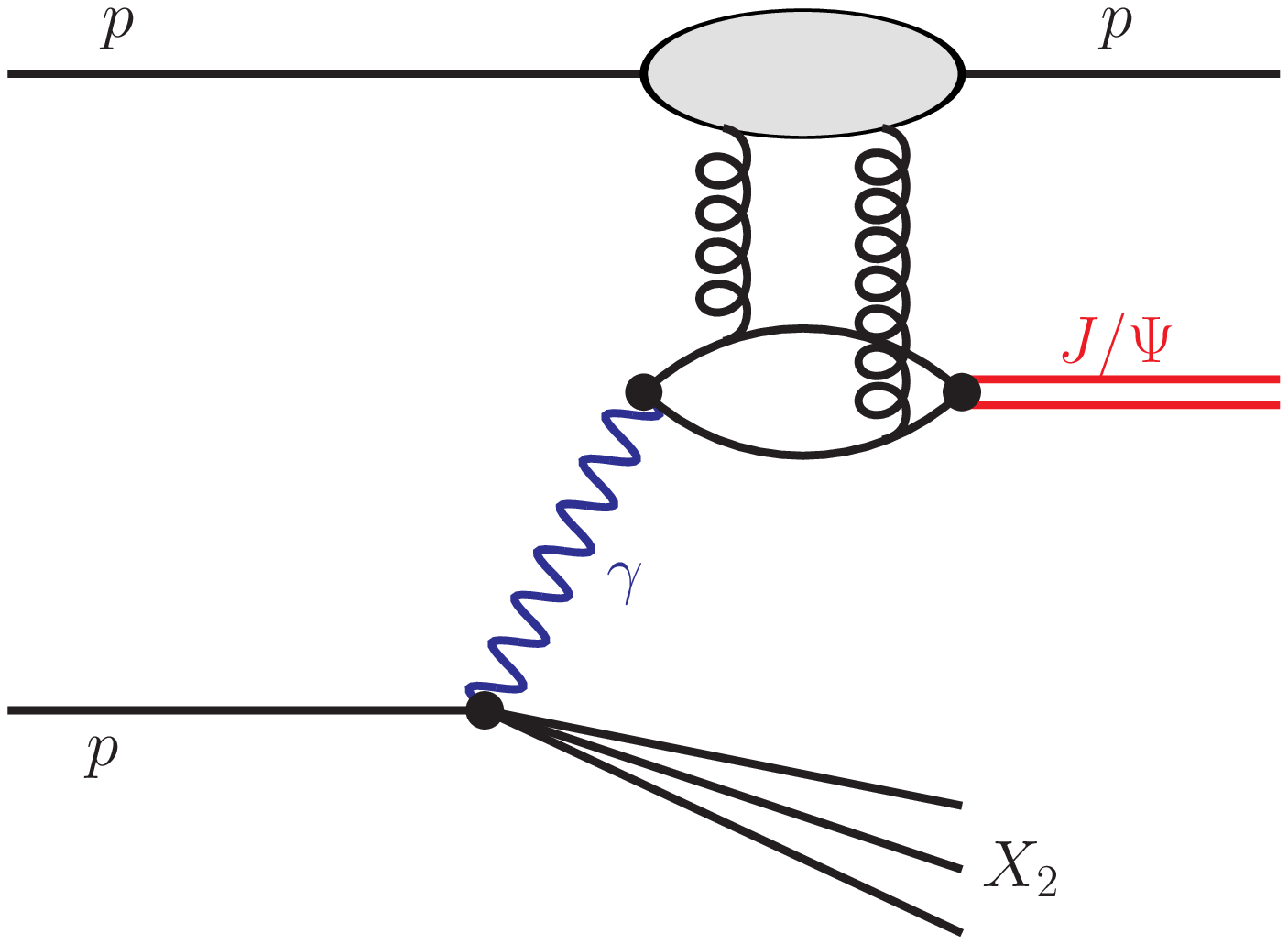}
\end{center}
\caption{Electromagnetic excitation of one of the photons
in proton-proton collisions.}
\label{fig:EM_dissociation}
\end{figure}

There are a few mechanisms of semiexclusive excitation of participating protons.
Let us start from electromagnetic dissociation of one of the protons
(see Fig.\ref{fig:EM_dissociation}).

The cross section for such processes can be written as:
\begin{eqnarray}
 {d \sigma (pp \to X V p; s) \over dy d^2\bp dM_X^2} = 
  \int {d^2\bq \over \pi \bq^2} {\cal{F}}^{(\mathrm{inel})}_{\gamma/p}(z_+,\bq^2,M_X^2) 
  {1\over \pi} {d \sigma^{\gamma^* p \to Vp} \over dt}(z_+s,t = -(\bq - \bp)^2) \nonumber \\
   +( z_+ \leftrightarrow z_-),  
  \nonumber \\
\end{eqnarray}
where $z_\pm = e^{\pm y} \sqrt{(\bp^2 + m_V^2)/s}$. 
In the kinematics of interest the ``fully unintegrated'' flux of photons associated with the breakup of the proton
is calculable in terms of the structure function $F_2$ of a proton \cite{daSilveira:2014jla,Luszczak:2015aoa}: 
\begin{eqnarray}
 {\cal{F}}^{(\mathrm{inel})}_{\gamma/p}(z,\bq^2,M_X^2) &=& {\alpha_{\mathrm{em}} \over \pi} (1 - z) \theta( M_X^2- M^2_{\mathrm{thr}})
 \nonumber \\
 &\times& {F_2(x_{Bj},Q^2)  \over M_X^2 + Q^2 - m_p^2}  \Big[ {\bq^2 \over \bq^2 + z (M_X^2 - m_p^2) + z^2 m_p^2} \Big]^2 
\, .
\nonumber \\
\end{eqnarray}

\begin{figure}
\begin{center}
\includegraphics[width=4.5cm]{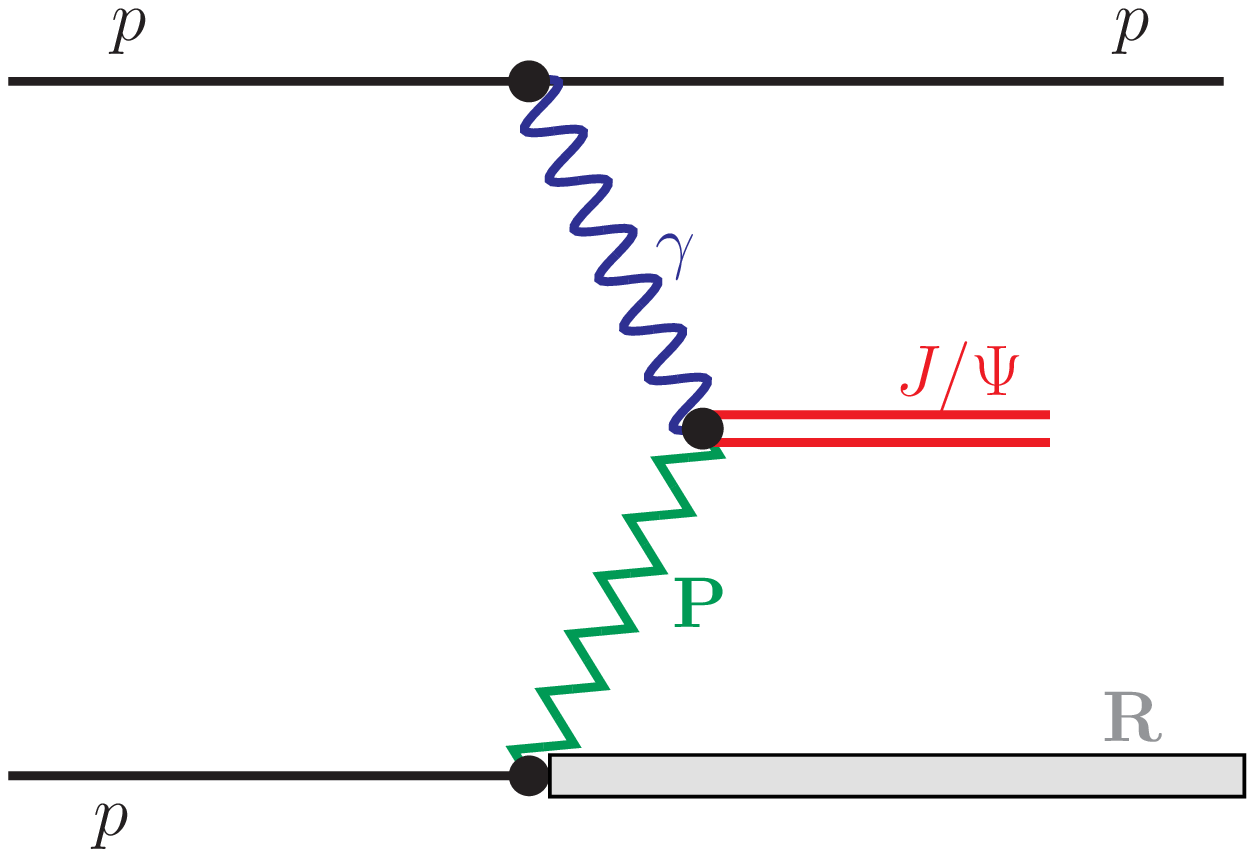}
\includegraphics[width=4.5cm]{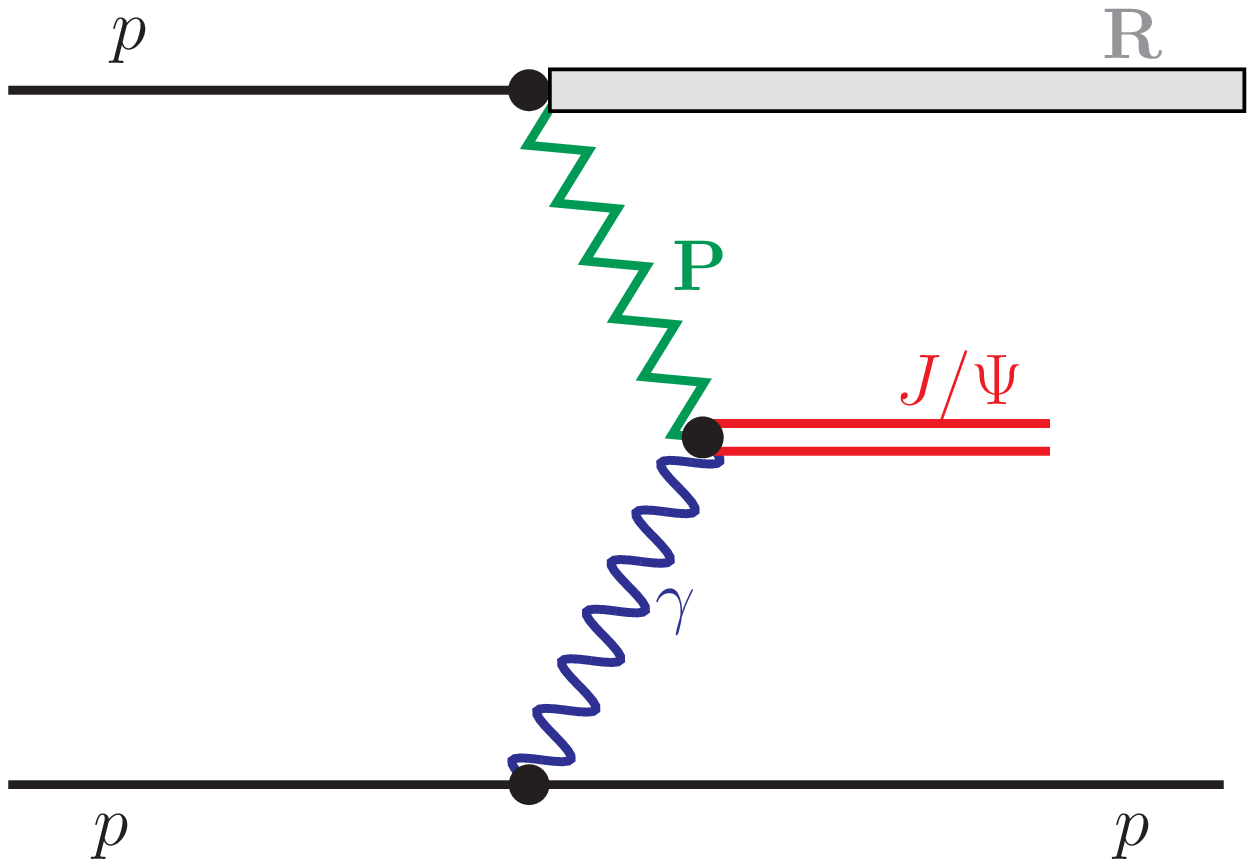}
\end{center}
\caption{Diffractive excitation of resonances on one of the protons.}
\label{fig:diff_res_dissociation}
\end{figure}

In distinction to electromagnetic dissociation, 
the diffractive excitation is highly model dependent.
We consider two mechanisms of the diffractive dissociation:
resonance production and continuum excitation.
A diagrammatic sketch of the reaction mechanism is shown
in Fig.\ref{fig:diff_res_dissociation}.

The contribution of three positive-parity baryon
resonances on the nucleon trajectory are taken into account: 
\begin{enumerate}
 \item N(1680), $J={5 \over 2}$,
 \item N(2220), $J={9 \over 2}$,
 \item N(2700), $J={13 \over 2}$.
\end{enumerate}
They contribute to the $p \Pom \to X$ amplitude as: 
\begin{eqnarray}
\Im m A(M_X^2,t) = \sum_{n=1,3} [f(t)]^{2(n+1)} \cdot { \Im m \, \alpha(M_X^2) \over 
(J_n - \Re e \, \alpha(M_X^2))^2 + (\Im m \, \alpha(M_X^2))^2} .
\end{eqnarray}
Here $J_n$ is the spin of the $n$th resonance, and the explicit form of 
the complex Regge trajectory $\alpha(M_X^2)$ as well as the form factor $f(t)$ 
(see \cite{JKLMO2011,Fiore:2004xb}).

In addition we include the so-called Roper resonance $N^*(1440)$.
More details are given in our original paper \cite{CSS2017}. 

\begin{figure}
\begin{center}
\includegraphics[width=4.5cm]{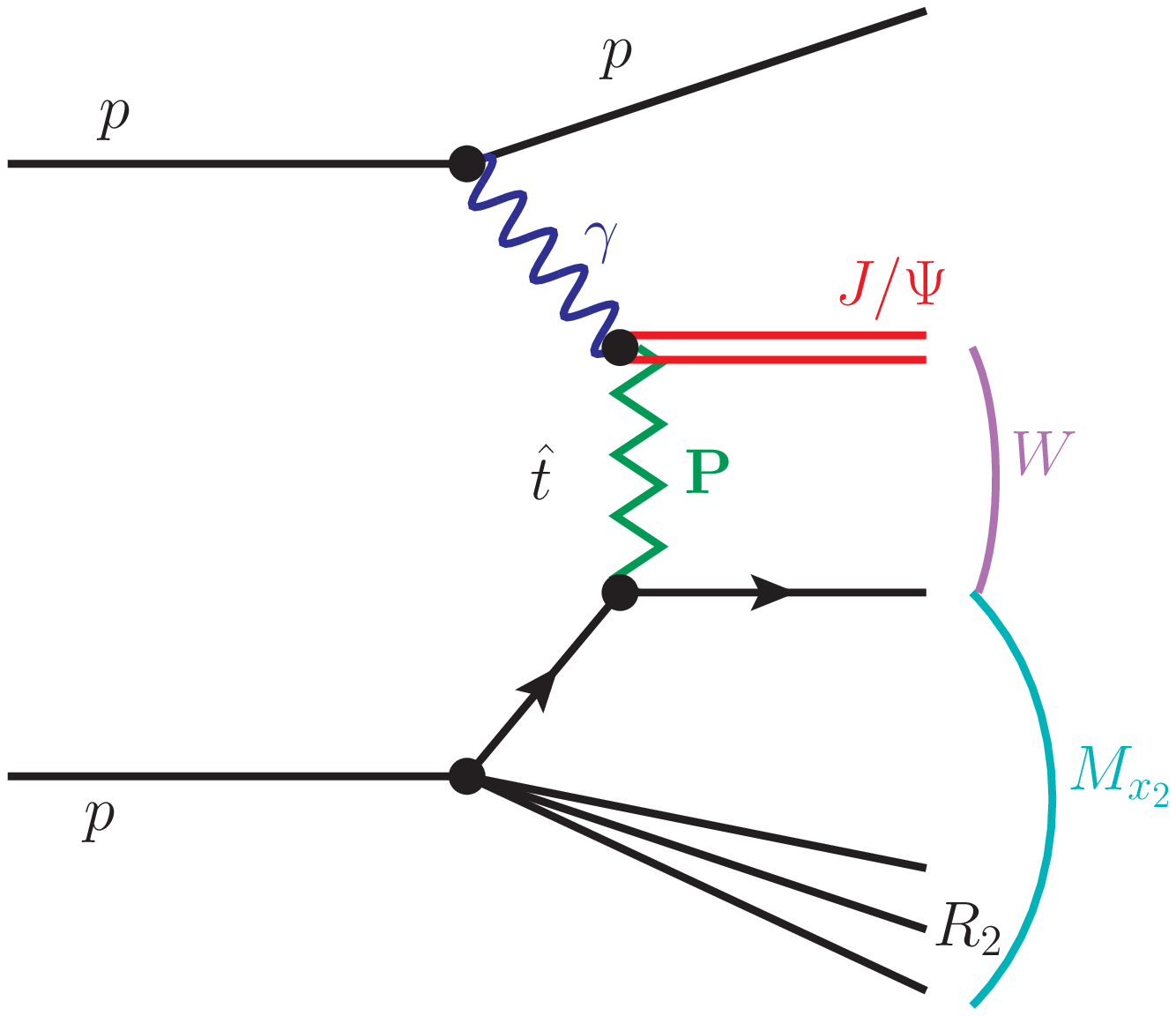}
\includegraphics[width=4.5cm]{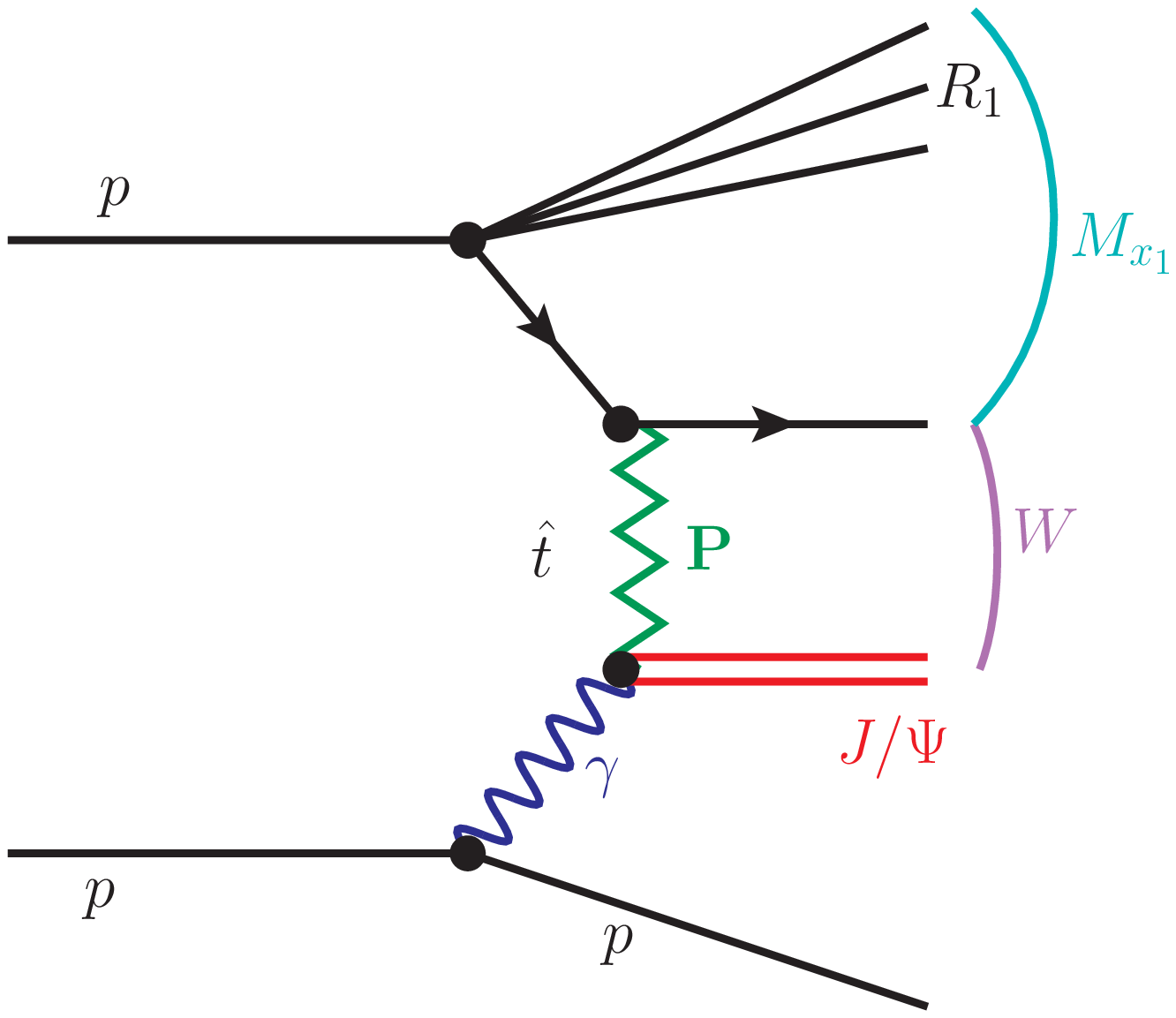}
\end{center}
\caption{Diffractive excitation of partonic continuum on one of 
the protons.}
\label{fig:diff_partonic_dissociation}
\end{figure}

We include also the partonic continuum for the $\gamma p \to V X$ subreaction.
Corresponding schematic diagrams are shown in 
Fig.\ref{fig:diff_partonic_dissociation}.

In this approximation the cross section can be written as
\begin{eqnarray}
\frac{d \sigma_{p p \to V j}^{diff,partonic}}{d y_V d y_j d^2 p_t}
&=& 
  \frac{1}{16 \pi^2 {\hat s}^2} 
x_1 q_{\rm eff}(x_1,\mu_F^2) x_2 \gamma_{el}(x_2)
\overline{| {\cal M}_{q \gamma \to V q} |^2}
\nonumber \\
&+& 
 \frac{1}{16 \pi^2 {\hat s}^2} 
 x_1 \gamma_{el}(x_1) x_2 q_{\rm eff}(x_2,\mu_F^2)
\overline{| {\cal M}_{q \gamma \to V q} |^2} \; .
\label{diffractive_partonic}
\end{eqnarray}
We neglect transverse momenta of the photon and the initial parton.
The two terms correspond to the two diagrams in 
Fig.\ref{fig:diff_partonic_dissociation}.
The effective parton distribution means:
\begin{equation}
q_{\rm eff}(x,\mu_F^2) = \frac{81}{16} g(x,\mu_F^2)
  + \sum_f \left[q_f(x,\mu_F^2) + {\bar q}_f(x,\mu_F^2) \right] \; .
\label{effective_quark_pdf}
\end{equation}
In Ref.\cite{CSS2017} we use a simple formula for two-gluon exchange:
\begin{equation}
\frac{d \sigma_{\gamma q \to V q}}{d \hat t} \propto \alpha_s^2({\bar Q}_t^2)
                                       \alpha_s^2(|\hat t|) 
           \frac{m_V^3 \Gamma(V \to l^+ l^-)}{({\bar Q}_t^2)^4} \; ,
\label{model_partonic_cs}
\end{equation}
where ${\bar Q}_t^2 = m_V^2 + |\hat t|$.
The normalization constant has been adjusted in \cite{CSS2017} 
to the H1 HERA experimental data.

\subsection{Selected results}

\begin{figure}
\begin{center}
\includegraphics[width=5.5cm]{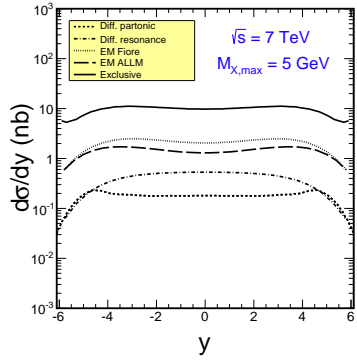}
\end{center}
\caption{Rapidity distribution of $J/\psi$ mesons for $\sqrt{s}$ = 7 TeV
for purely exclusive process and for different contributions 
to semiexclusive processes.}
\label{fig:dissociation_LHCb_y}
\end{figure}

In Ref.\cite{CSS2017} we have shown many results for both full phase
space as well as for particular experiments.
In Fig.\ref{fig:dissociation_LHCb_y} we show the distribution in rapidity
of the $J/\psi$ meson.

\begin{figure}
\begin{center}
\includegraphics[width=5.5cm]{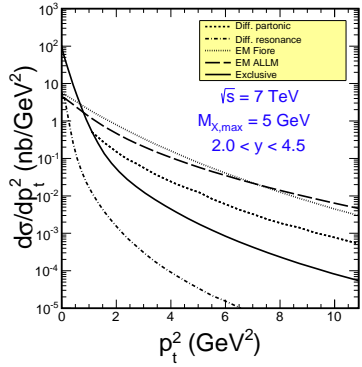}
\end{center}
\caption{Dependence of the cross section on transverse momentum
squared for the LHCb experiment.}
\label{fig:dissociation_LHCb_pt2}
\end{figure}

In Fig.\ref{fig:dissociation_LHCb_pt2} we show 
transverse momentum distribution of $J/\psi$ meson for 
$M_X < 5 ~\rm{GeV}$.
The larger masses of the proton excitations the larger slopes 
of the transverse momentum distribution of $J/\psi$.
For comparison we show distribution for the purely exclusive case.
The inelastic contributions start to dominate over the purely
elastic (exclusive) contribution for $p_T > 1 ~\rm{GeV}$.

\begin{figure}
\begin{center}
\includegraphics[width=5.5cm]{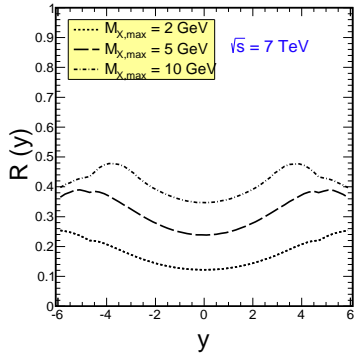}
\end{center}
\caption{The ratio of the semiexclusive-to-exclusive cross section
as a function of rapidity of $J/\psi$ meson for different ranges
of proton excitation energies.}
\label{fig:dissociation_ratio}
\end{figure}

It is also interesting to see how the cross sections for semiexclusive
processes compare to those for the purely exclusive one.
To quantify their contribution we define the following ratio:
\begin{eqnarray}
R(y) &=& \frac{d\sigma_{p p \to p J/\psi X}(M_X < M_{X,\rm{max}})/dy}
              {d\sigma_{p p \to p J/\psi p}/dy} \; . 
\label{ratio_inelastic_to_elastic} 
\end{eqnarray}
This ratio is shown in Fig.\ref{fig:dissociation_ratio} for 
three different values of $M_{X,\rm{max}}$.
We see that the magnitude as well as the shape of the ratio depends 
on the range of missing masses included in the calculation.

\section{Inclusive production of $J/\psi$ mesons}

\subsection{Sketch of the theoretical methods}
\begin{figure}
\begin{center}
\includegraphics[width=4.5cm]{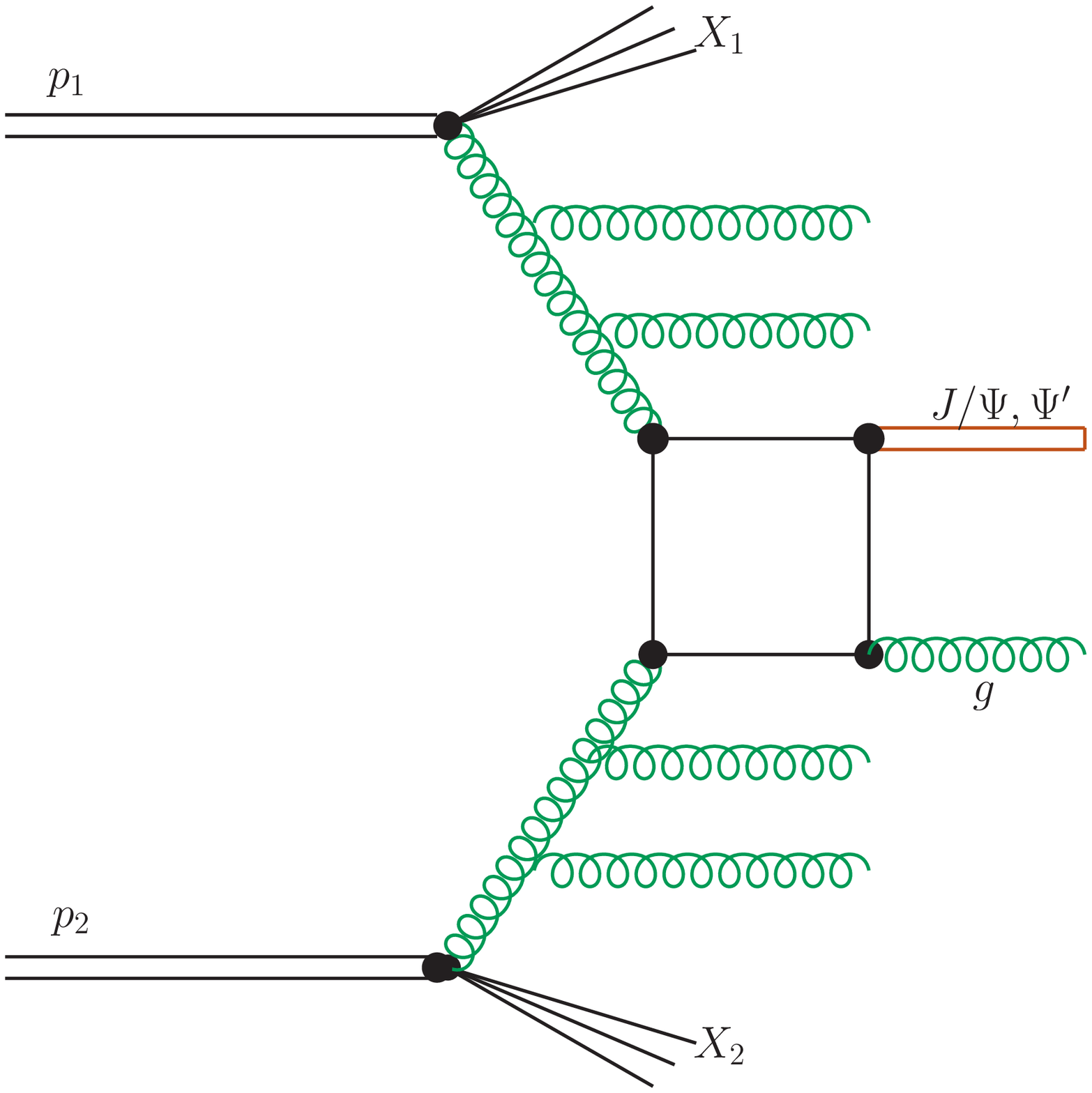}
\includegraphics[width=4.5cm]{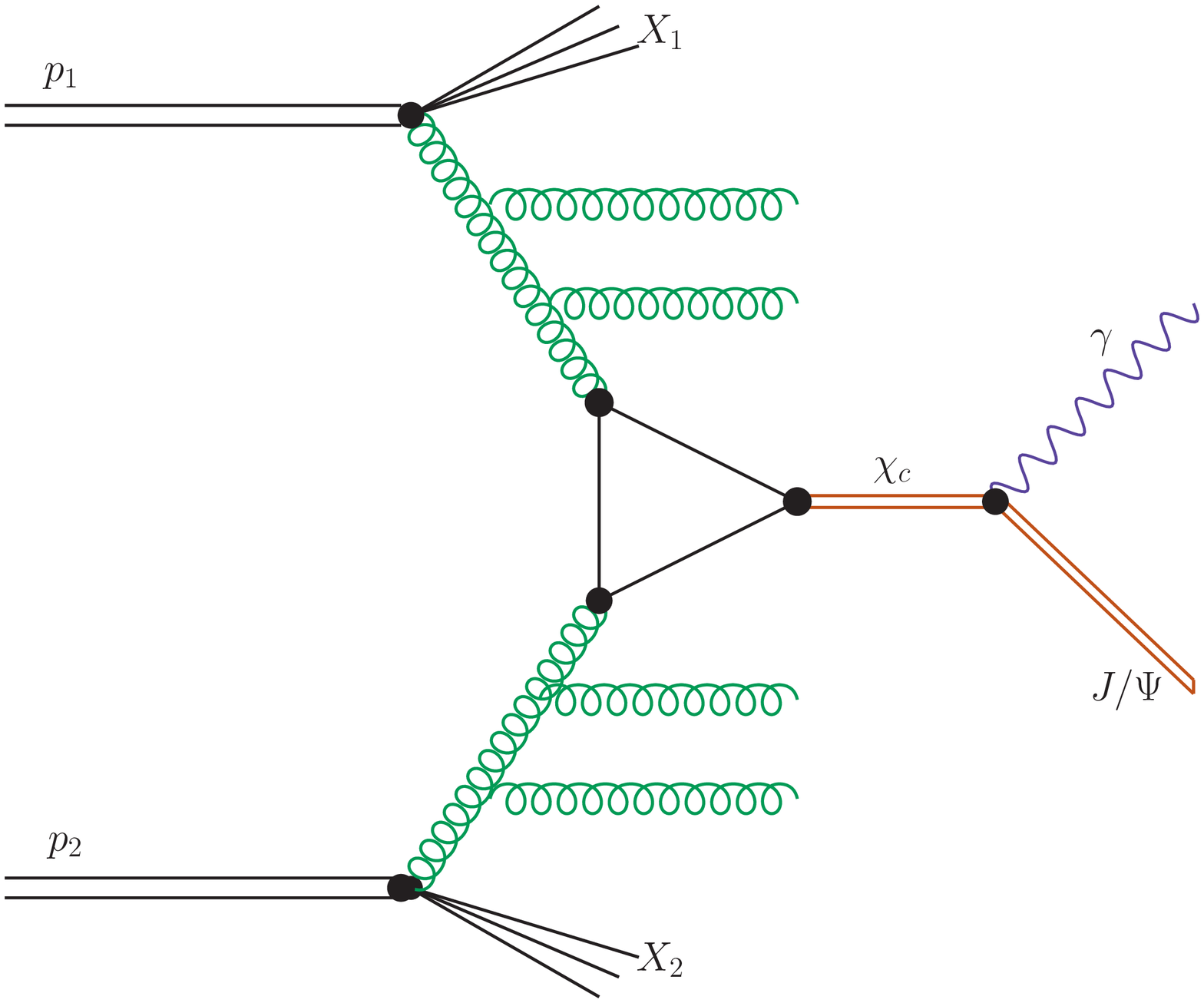}
\end{center}
\caption{Dominant color-singlet mechanisms of $J/\psi$ meson
  production. The initial state gluon emissions are shown explicitly.}
\label{fig:inclusive_jpsi}
\end{figure}

We now turn to the fully inclusive production of $J/\psi$ mesons.
In this presentation we shall consider only the contributions of the 
color-singlet mechanism. Here at the perturbative stage of the process
a $c \bar c$ system is produced in the color-singlet state.
In Figs.\ref{fig:inclusive_jpsi} we show two dominant color-singlet 
mechanisms of $J/\psi$ meson production.
The first diagram represents so-called direct contribution, while the
second diagram represents an example of feed down from other quarkonia.

%
\begin{eqnarray}
&&\frac{d \sigma(p p \to J/\psi g X)}{d y_{J/\psi} d y_g d^2 p_{J/\psi,t} d^2 p_{g,t}}
 = 
\frac{1}{16 \pi^2 {\hat s}^2} \int \frac{d^2 q_{1t}}{\pi} \frac{d^2 q_{2t}}{\pi} 
\overline{|{\cal M}_{g^{*} g^{*} \rightarrow J/\psi g}^{off-shell}|^2}  
\nonumber \\
&& \times \;\; 
\delta^2 \left( \vec{q}_{1t} + \vec{q}_{2t} - \vec{p}_{H,t} - \vec{p}_{g,t} \right)
{\cal F}_g(x_1,q_{1t}^2,\mu_{F}^{2}) 
{\cal F}_g(x_2,q_{2t}^2,\mu_{F}^{2}) \; .
\label{kt_fact_gg_jpsig}
\end{eqnarray}
The corresponding matrix element squared for the $g g \to J/\psi g$ is proportional
to the radial wavefunction at the origin:
\begin{equation}
|{\cal M}_{gg \to J/\psi g}|^2 \propto \alpha_s^3 |R(0)|^2 \; .
\label{matrix_element} 
\end{equation}
In our calculations the matrix elements obtained by S. Baranov 
\cite{Baranov_private} was used.

Another important mechanism is the production of $\chi_c$ mesons
and its subsequent decay $\chi_c \to J/\psi + \gamma$.
The $\chi_c$ mesons are $p$-wave $c \bar c$ states of positive
$C$-parity. They can be produced by the gluon-gluon fusion mechanism.
In the $k_t$-factorization approach the leading-order cross section for 
the $\chi_c$ meson production can be written in a similar way as the 
for the $J/\psi$ production.

The matrix element squared for the $g^* g^* \to \chi_c$ subprocess is
proportional to the derivative of the wave function at the origin:
\begin{equation}
|{\cal M}_{g^*g^* \to \chi_c}|^2 \propto \alpha_s^2 |R'(0)|^2 \; .
\label{matrix_element} 
\end{equation}
$R'(0)$ can be treated as a free parameter to get correctly
relative contribution of $J/\psi$ from $\chi_c$. 
We used the matrix element taken from \cite{KVS2006}. 

Another mechanism which we take into account is the production
of $\psi'$ meson and its decay. The corresponding ratio
$\rm{Br}(\Psi' \to J/\psi + X) =$ 0.574 is relatively large.
The cross section for production of $\psi'$ mesons is calculated 
using a formula analogous to Eq.(\ref{kt_fact_gg_jpsig}).
Of course, then the wave function of $J/\psi$ is replaced
by a wave function of $\psi'$:
$|R_{\Psi'}(0)|^2 \approx 5/8 |R_{J/\psi}(0)|^2$.
The difference may be understood qualitatively as due to the fact that
$\psi'$ is a 2S state while $J/\psi$ is 1S state (different radial 
excitations).

\subsection{Selected results}

\begin{figure}
\begin{center}
\includegraphics[width=4cm]{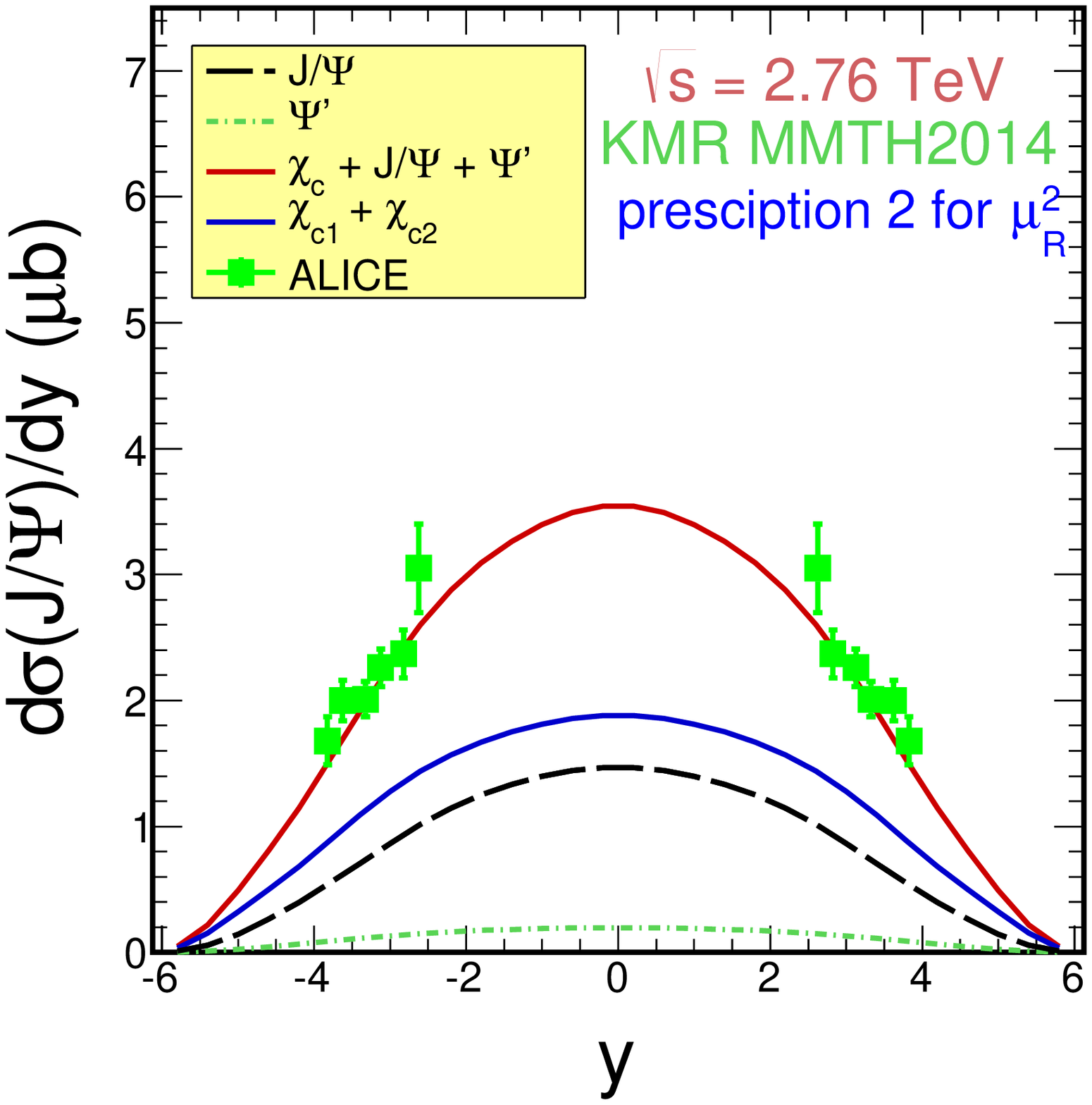}
\includegraphics[width=4cm]{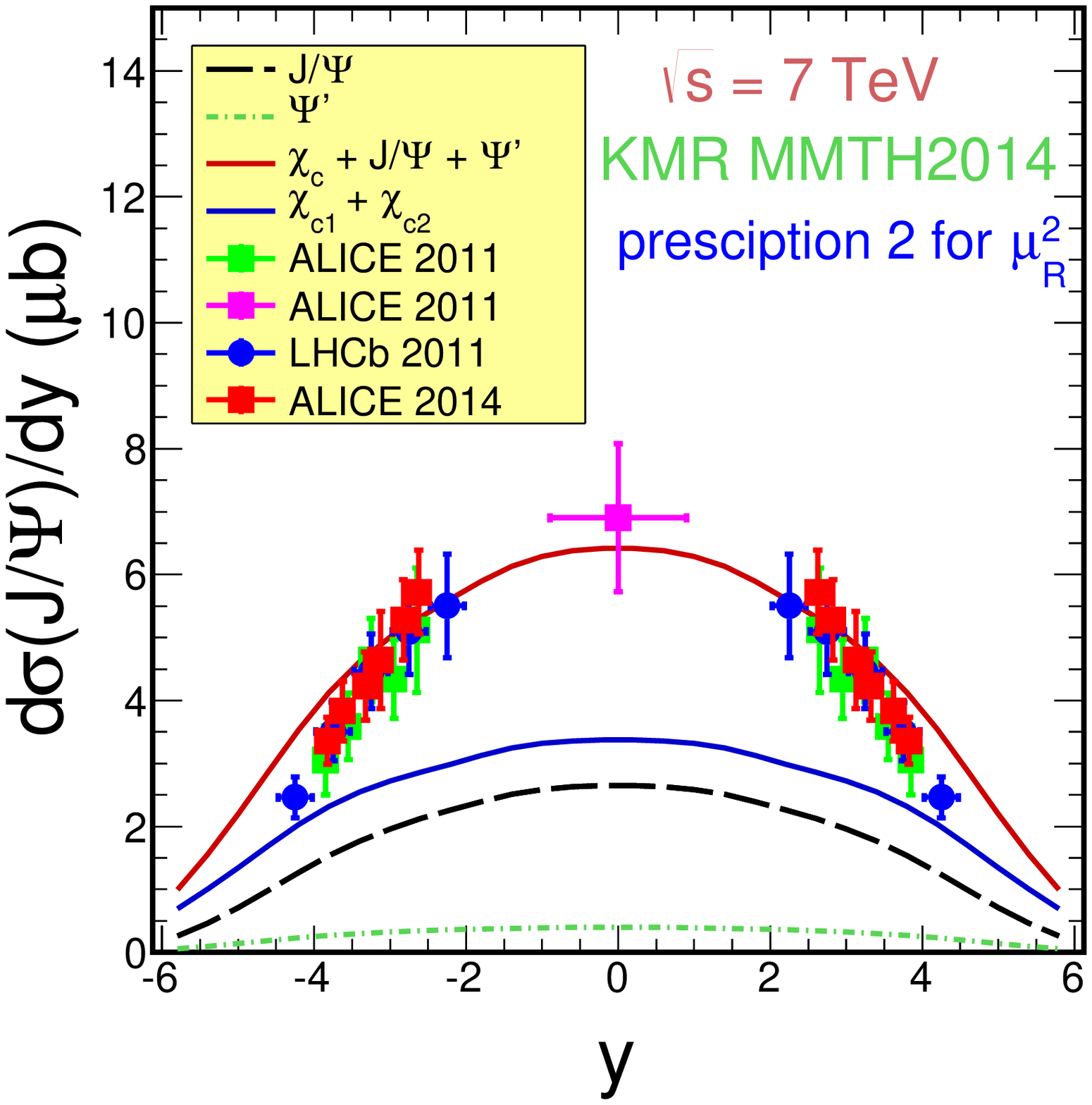}
\includegraphics[width=4cm]{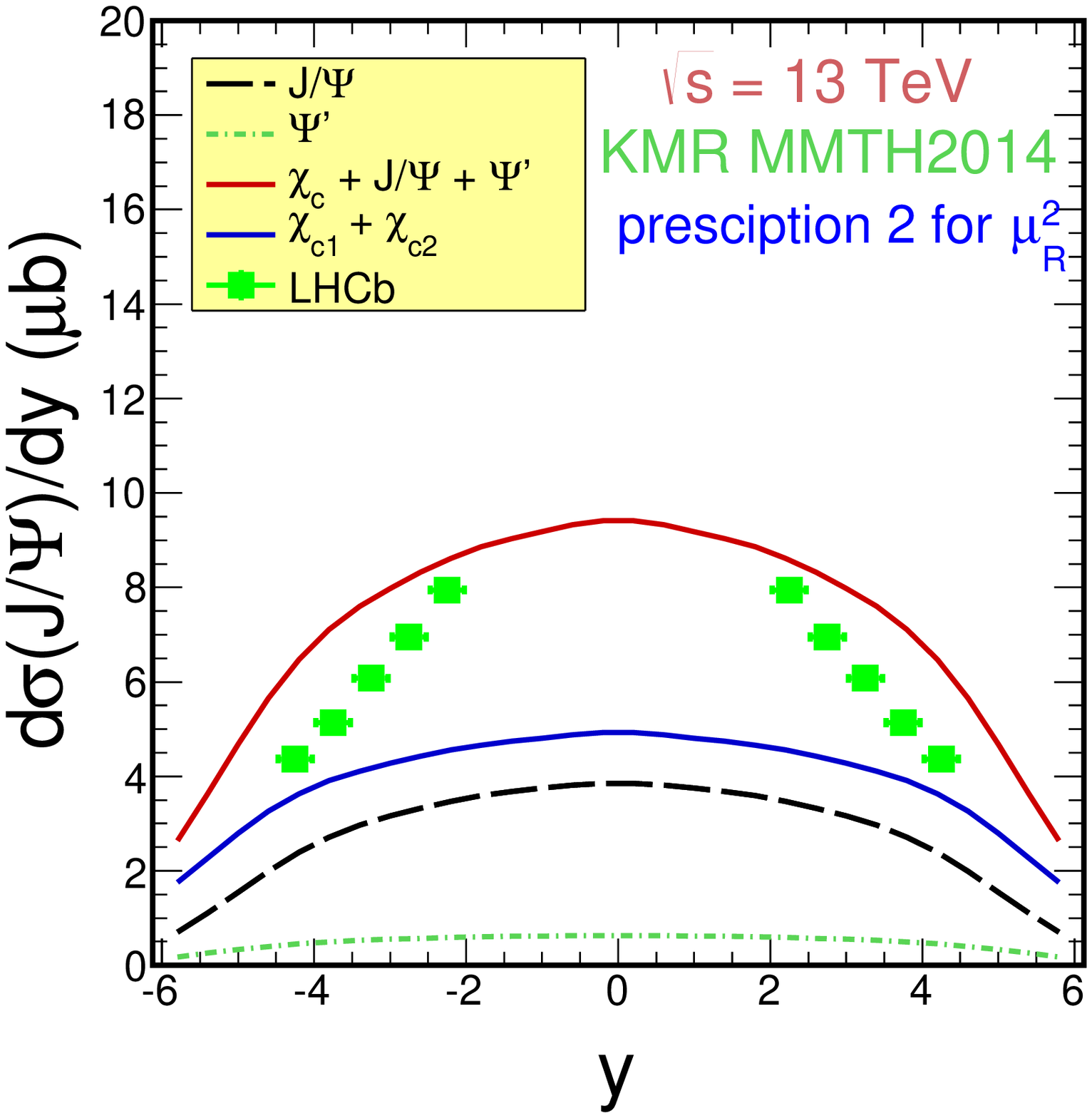}
\end{center}
\caption{Inclusive rapidity distributions of $J/\psi$ mesons
for three different energies obtained with the KMR UGDFs.}
\label{fig:dsig_dy_KMR}
\end{figure}

Here we show only some selected results. More detailed discussion
will be presented elsewhere \cite{CS2017}.
In Fig.\ref{fig:dsig_dy_KMR} we show rapidity distributions
obtained with the KMR UGDF. This gluon distribution gives
a relatively good description of the open charm inclusive production
and correlations. At low energies relatively good agreement is obtained.
When increasing energy the agreement becomes worse and worse.
The disagreement may be connected with the onset of nonlinear
effects, leading eventually to gluon saturation which is an asymptotic
(small-$x$) notion.

For the LHCb configuration typically one longitudinal momentum
fraction is small and the other much larger. Replacing the KMR UGDF by 
the Kutak-Stasto UGDF for the low-$x$ values improves the situation. 
In our opinion it seems precocious to draw too definite conclusion 
in the moment.

\begin{figure}
\begin{center}
\includegraphics[width=4cm]{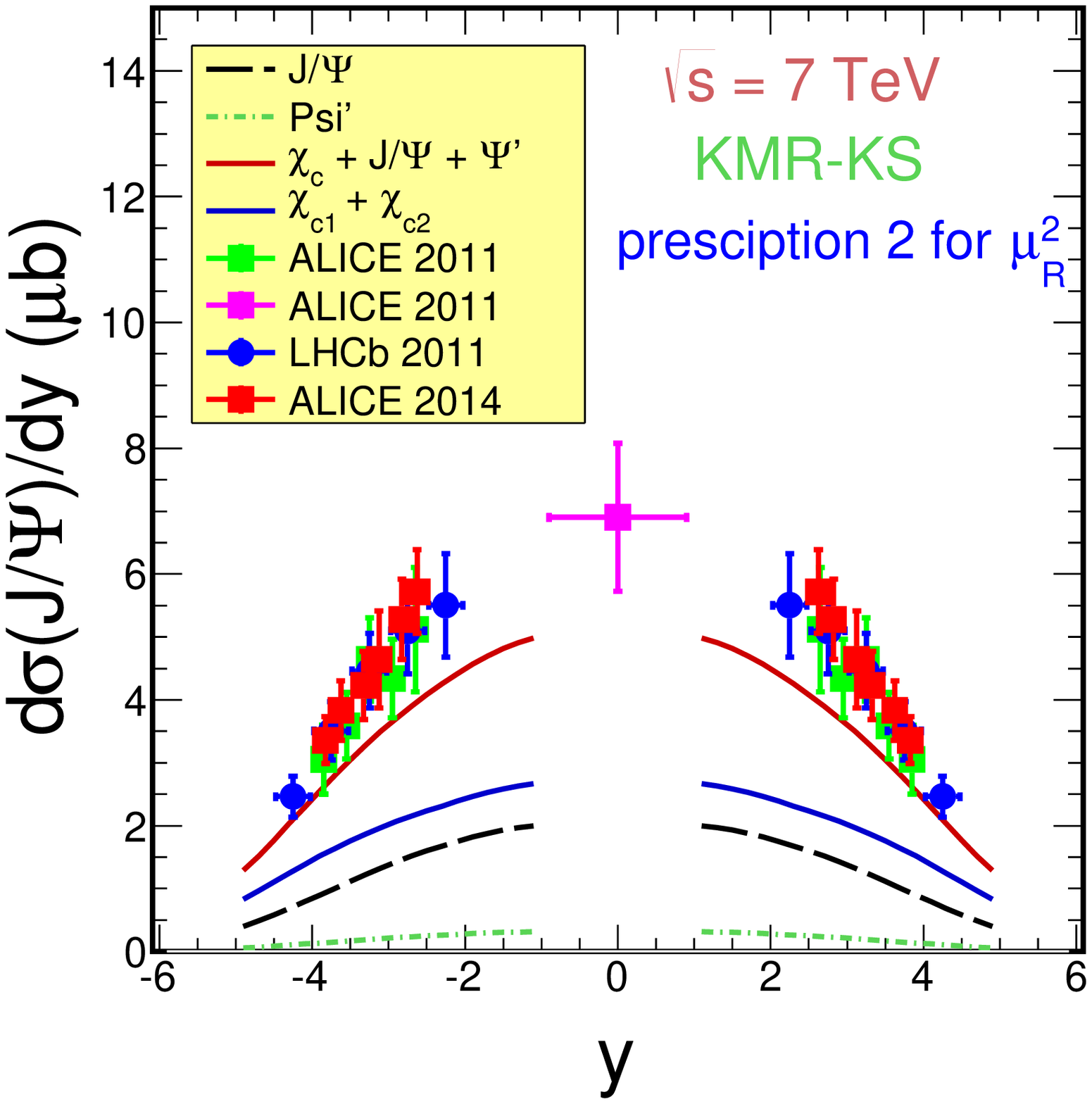}
\includegraphics[width=4cm]{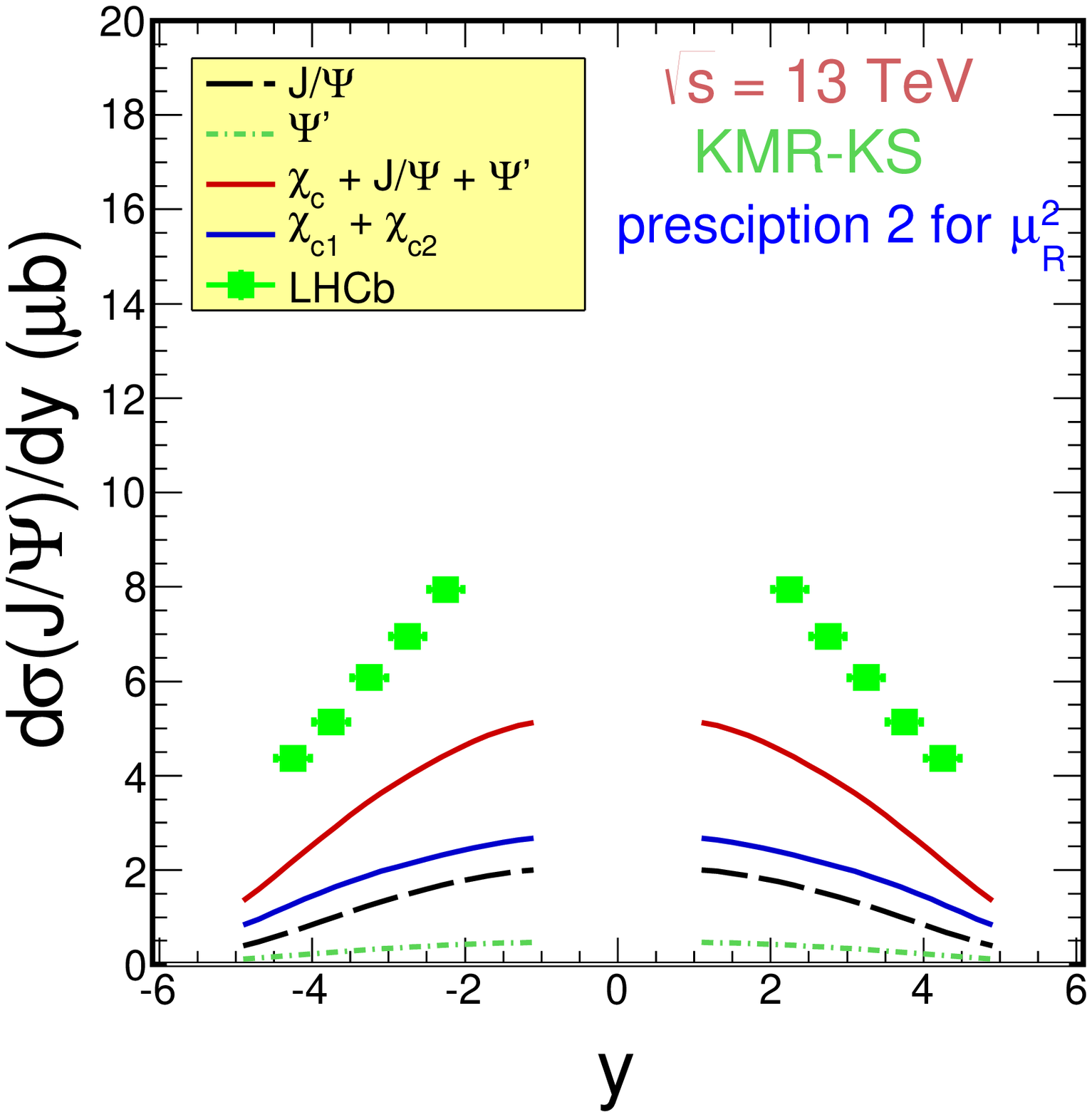}
\end{center}
\caption{Inclusive rapidity distributions of $J/\psi$ mesons
for two energies obtained with one KMR and one KS(nonlinear) UGDF.}
\label{fig:dsig_dy_mixed}
\end{figure}

In \cite{CS2017} we will discuss many other aspects of $J/\psi$ meson
production.

\section{Production of two $J/\psi$ mesons}

\subsection{Sketch of the theoretical methods}

\begin{figure}
\begin{center}
\includegraphics[width=5cm]{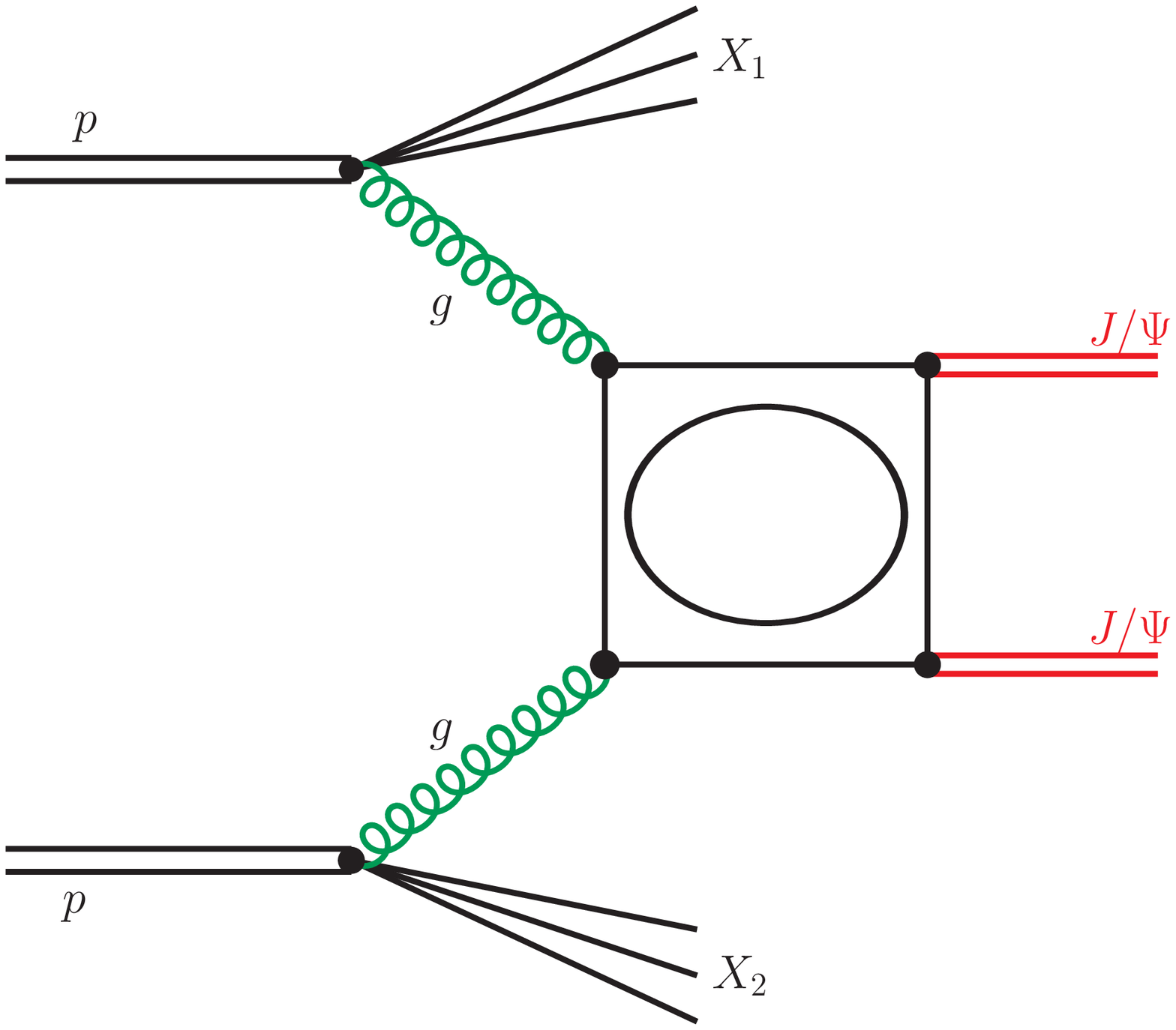}
\includegraphics[width=5cm]{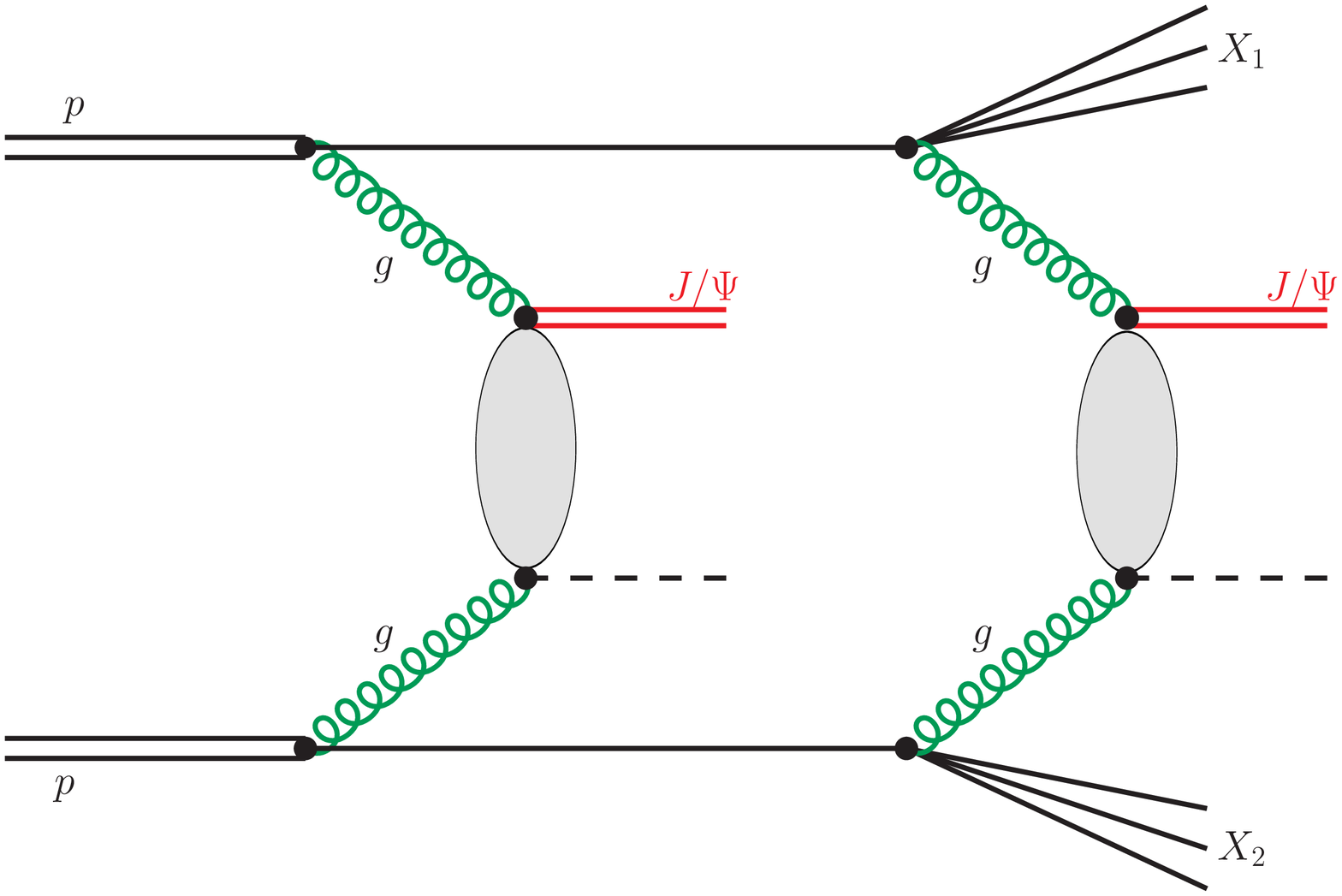}
\end{center}
\caption{Schematic representation of main mechanisms considered so
  far in the literature.}
\label{fig:jpsijpsi_standard} 
\end{figure}

The color-singlet mechanisms used so far in the literature are shown
in Fig.\ref{fig:jpsijpsi_standard}.
The first type of diagrams will be called here ``box'' for brevity.
The circle inside the box represents different gluon insertions.
There are 20 diagrams (see e.g.\cite{BSZSS2013}).
These contributions are of the order of
$O(\alpha_s^4)$. In addition they depend
on the value of the wave function at the origin 
$\sigma \propto |R(0)|^4$. 
The matrix elements are too complicated to be shown here
explicitly. The second diagram, of the order of $O(\alpha_s^6)$
represents double parton scattering mechanism. Although higher
order than the previous box contributions it may be enhanced by higher
powers of gluon distributions.

For the presentation at the Epiphany 2017 workshop
the box contribution was calculated only in the collinear approach. 
\footnote{Now a full $k_t$-factorization approach is available but will
  be not shown here.}
Then the inclusive differential cross section can be written as: 
\begin{eqnarray}
&&\frac{d \sigma(p p \to J/\psi J/\psi}
{d y_{V_1} d y_{V_2} d^2 p_t}
 = 
\frac{1}{16 \pi^2 {\hat s}^2}  
\overline{|{\cal M}_{g g \rightarrow J/\psi J/\psi}^{on-shell}|^2} 
\nonumber \\
&& \times \;\; 
g(x_1,\mu_{F}^{2}) g(x_2,\mu_{F}^{2}) \; .
\label{collinear_gg_jpsijpsi}
\end{eqnarray}
In our calculations for the presentation we used the MSTW08 
gluon distributions. We have checked that the somewhat artificial
matrix element gives a good representation of experimental data
for inclusive $J/\psi$ production.

The calculation of double parton scattering is simplified.
Instead of including all single scattering mechanisms discussed in the
previous section we parametrize the single $J/\psi$ production using
the following formula:
\begin{equation}
\frac{d \sigma(p p \to J/\psi g)}{d y_{J/\psi} d y_{X} d^2 p_t}
 = 
\frac{1}{16 \pi^2 {\hat s}^2}  
\overline{|{\cal M}_{g g \rightarrow J/\psi X}^{eff}|^2}  
\times \;\; 
g(x_1,\mu_{F}^{2}) g(x_2,\mu_{F}^{2}) \;  .
\label{effective_ME}
\end{equation}
We take a parametrization of the effective matrix element 
by Kom-Kulesza-Stirling \cite{KKS2011}
using the MSTW08 PDF.

In the present approach we do calculations using a 
factorized Ansatz. Then the cross section for 
double parton scattering can be written as:
\begin{equation}
\frac{d \sigma}{d y_1 d^2 p_{1t} d y_2 d^2 p_{2t}} 
= \frac{1}{2 \sigma_{eff}} \cdot
  \frac{d \sigma}{d y_1 d^{2}p_{1t}} \cdot
  \frac{d \sigma}{d y_2 d^{2}p_{2t}}  \;  .
\end{equation}

The $\sigma_{eff}$ parameter is in principle a free parameter
responsible for the overlap of partonic densities of colliding protons.
$\sigma_{eff}$ = 15 mb is world average for different reactions.
We shall use this value having in mind that a departure from the
factorized Ansatz is possible. A complete calculation of the single
parton scattering contributions should show how much room is left
for the DPS contribution, which by itself is rather difficult 
to be calculated from first principle.

\begin{figure}
\begin{center}
\includegraphics[width=5cm]{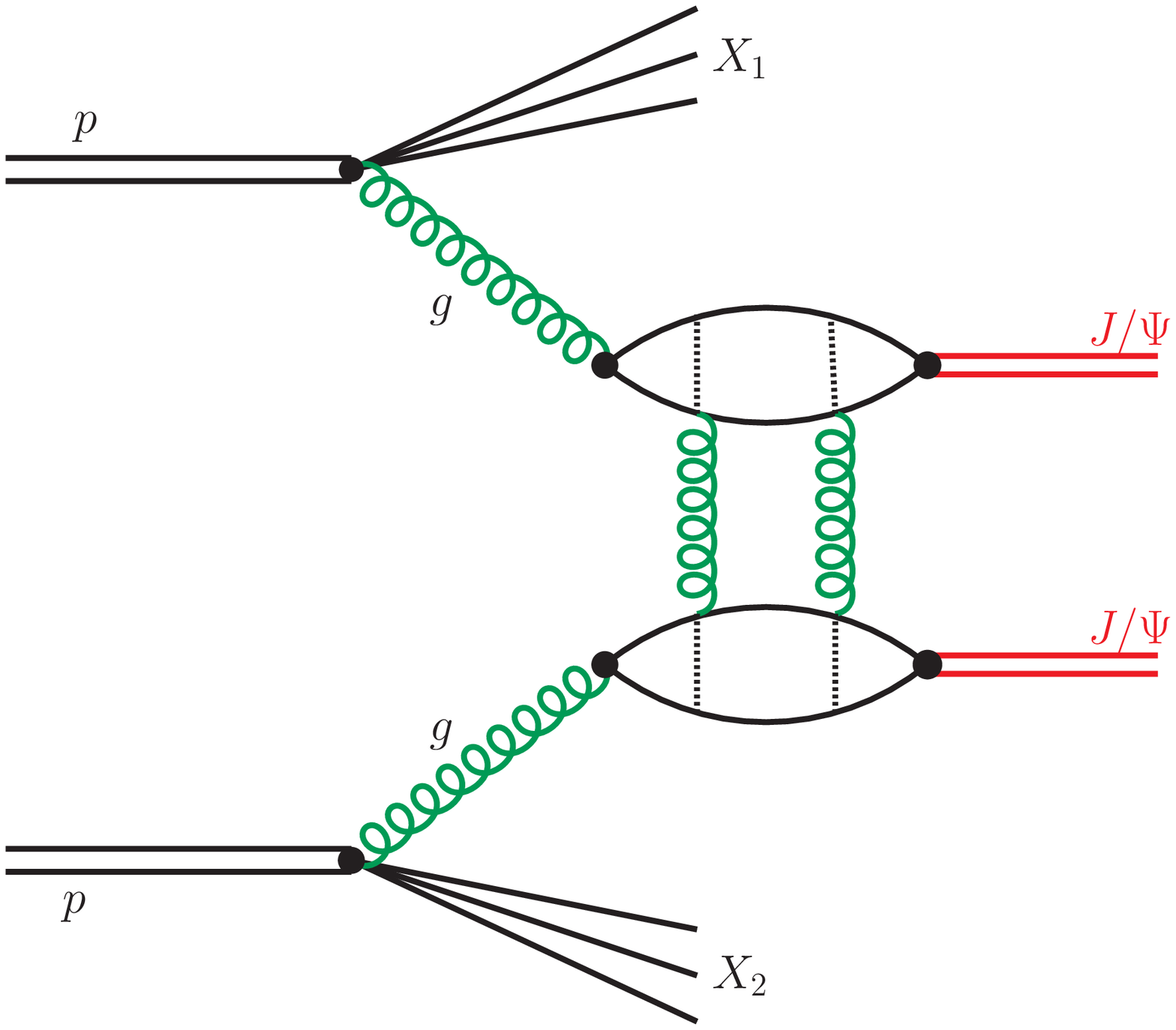}
\includegraphics[width=5cm]{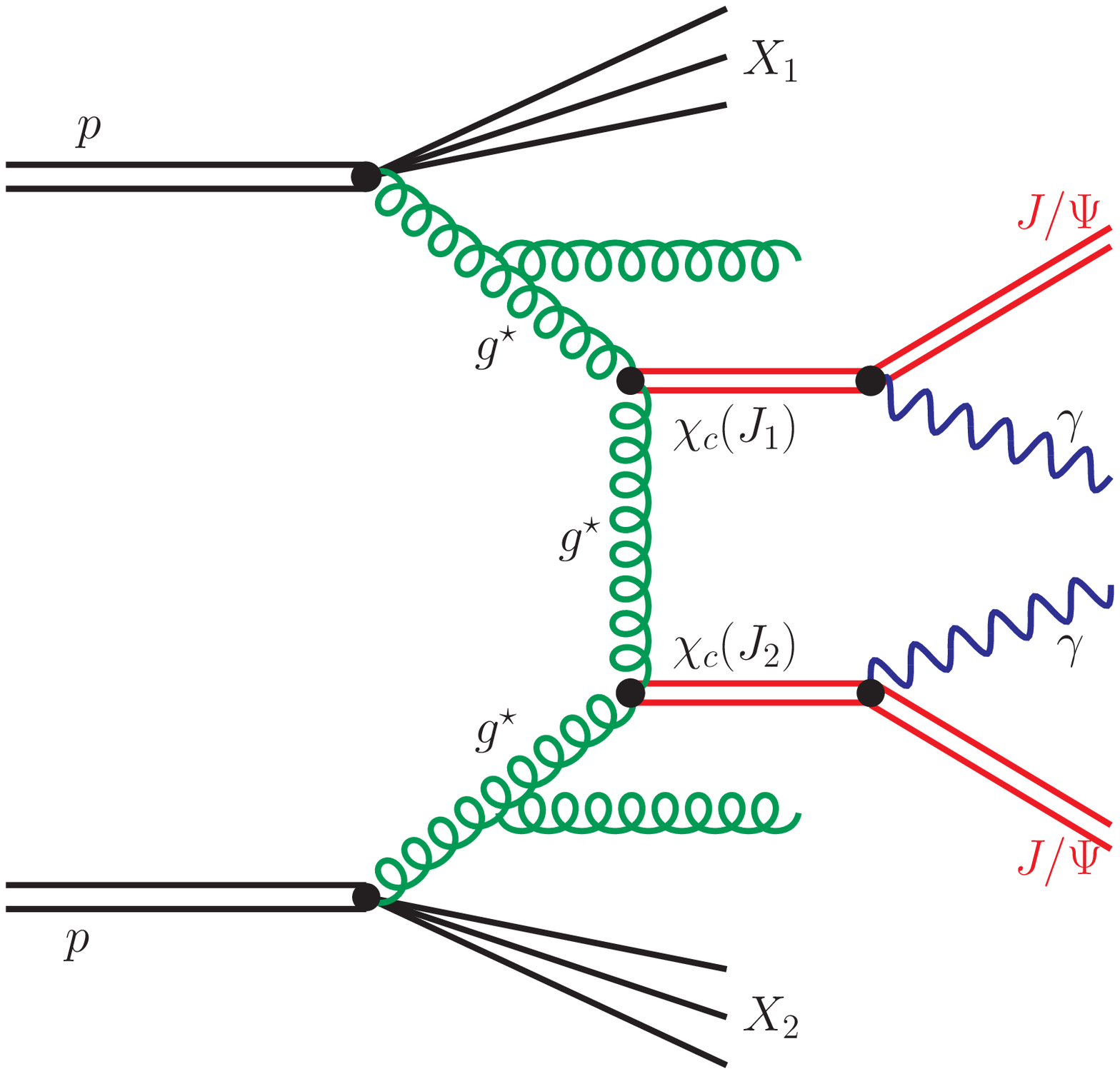}
\end{center}
\caption{New mechanisms included or being included by us recently.}
\label{fig:jpsijpsi_new}
\end{figure}

In Fig.\ref{fig:jpsijpsi_new} we show mechanisms not included routinely
in the calculation of simultaneous two $J/\psi$ meson production.
The first one (16 diagrams) was included before in \cite{BSZSS2012}
but was there not crucial. The second one was not discussed in
the presentation at Epiphany, but will be presented soon elsewhere.

\subsection{Selected results}

Here we show some preliminary results obtained for very recent
ATLAS data \cite{ATLAS_jpsijpsi}. Both cuts on $J/\psi$ and
muon kinematical variables have been imposed. The details will
be given elsewhere \cite{CSS2017}.

\begin{figure}
\begin{center}
\includegraphics[width=7cm]{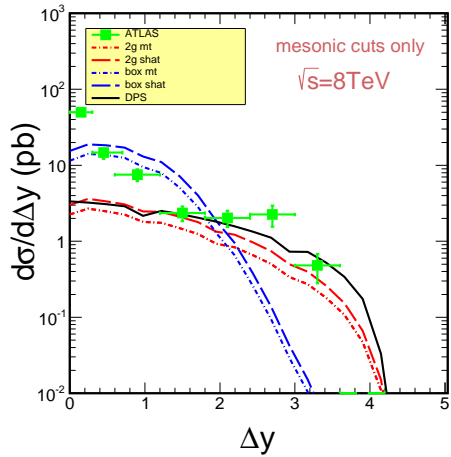}
\end{center}
\caption{Distribution in rapidity difference between the two $J/\psi$ mesons.
Contributions of different mechanisms are shown separately.}
\label{fig:jpsijpsi_dsig_dydiff}
\end{figure}

In Fig.\ref{fig:jpsijpsi_dsig_dydiff} we present distribution
in rapidity difference between the two $J/\psi$ mesons.
Both the leading-order and two-gluon exchange contributions are
shown for two different factorization scales.
Summing the different contributions one can almost describe
the experimental data except of very small rapidity differences. 
We ask the reader to note a similar shape of distributions 
for DPS and two-gluon exchange.

\begin{figure}
\begin{center}
\includegraphics[width=7cm]{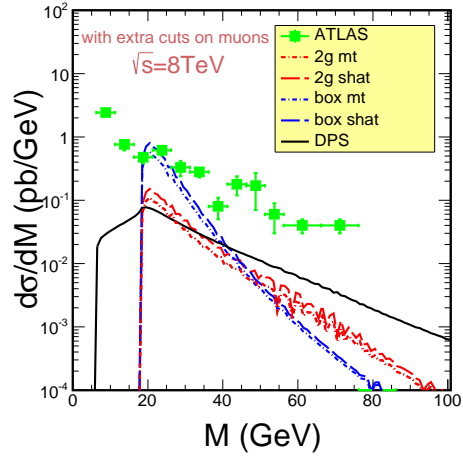}
\end{center}
\caption{Invariant mass distribution of two $J/\psi$ mesons.
Contributions of different mechanisms are shown separately.}
\label{fig:jpsijpsi_dsig_dMVV}
\end{figure}

In Fig.\ref{fig:jpsijpsi_dsig_dMVV} we show distribution in dimeson
invariant mass. In collinear-factorization approach and with the ATLAS 
experimental cuts on transverse momenta of $J/\psi$ mesons there is a sharp
cut off in meson invariant mass ($M >$ 20 GeV).
There is no such a sharp cut off for the DPS contributions where both $J/\psi$
mesons are not correlated in azimuthal angle.
Although we do not show the sum of the contributions it is obvious that
no good description of the data is possible, especially for low
invariant masses. This is will be discussed in detail in our
forthcoming paper \cite{CSS2017}.

\begin{figure}
\begin{center}
\includegraphics[width=7cm]{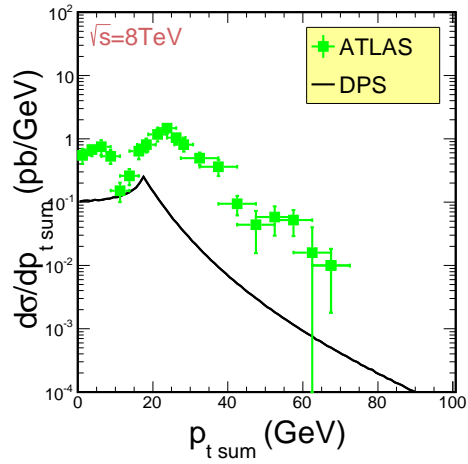}
\end{center}
\caption{Distribution of $p_{t,sum}$. Only the DPS contribution
is shown and compared to the ATLAS data.}
\label{fig:jpsijpsi_dsig_dptsum}
\end{figure}

In collinear-factorization approach only $d \sigma/d y_{diff}$ and
$d \sigma/ dM_{VV}$ can be obtained.
For the DPS mechanism also $p_{t,sum} = |{\vec p}_{1,t} + {\vec p}_{2,t}|$
distribution can be obtained as shown in 
Fig.\ref{fig:jpsijpsi_dsig_dptsum}. Clearly the DPS contribution is
not sufficient as can be seen by comparison of the cross section
normalizations and in particular of the shapes of the theoretical and
experimental distributions.

\section{Conclusions}

In the present paper we have discussed different reactions with 
the $J/\psi$ meson in the final state.

We have started from exclusive $p p \to p p J/\psi$ process.
In contrast to other groups we include interference effect between
the photon-pomeron and pomeron-photon amplitudes 
and present distributions in $J/\psi$ transverse momentum.
We have quantified the effect of including the anomalous photon coupling to
nucleon and those due to absorption effects, both often neglected
in the literature. Our calculation suggest that the LHCb data
show an onset of nonlinear effects.

However, some caution is required when interpreting the present
experimental results. We have discussed that the present measurements
were not fully exclusive as protons were not measured and only rapidity
gap conditions were imposed experimentally. We have discussed the
reaction in which one of the protons is dissociated, but large 
rapidity gaps around $J/\psi$ are present. 
We called them semi-exclusive reactions. 
In general, there are two classes of the dissociative processes: 
associated with photon exchange and associated with pomeron exchange.
We have presented a novel method to include electromagnetic
dissociation. In this method the $p \to X$ vertex is expressed in
terms of $F_2$ structure function. By using parametrizations valid
in the full phase space one can include both resonance and continuum
excitations. Diffractive excitations are more difficult to model.
We have discussed both resonance contributions using a model from
the literature and partonic continuum caused by the exchange of colour
singlet two-gluon exchange with parameters adjusted to HERA data.
We have shown that, in disagreement with naive expectation in the field,
the electromagnetic excitation gives larger contribution than the
diffractive one. The effect of the semiexclusive processes
on rapidity and transverse momenta of the $J/\psi$ mesons have been
presented.

Paradoxically the inclusive production of $J/\psi$ mesons is less
understood than the exclusive one. Different authors have different
opinions on the underlying mechanism. The difficulty is that the
higher-order effects are more important than for other pQCD processes.
In the present short review we have considered only so-called 
color-singlet mechanisms that, in contrast to color-octet mechanisms,  
can be calculated from first principles. To include higher order
effects we have used the $k_t$-factorization method and modern
unintegrated gluon distributions. In this presentation we have 
focused on the description of recent LHCb data for forwardly produced
$J/\psi$ mesons. Both direct and feed-down contributions were
included. The rather forward production, especially for $\chi_c$
feed-down, is sensitive to small values of longitudinal momentum
fraction. We have obtained a good description of lower energy rapidity
distributions and gradually worse and worse description with
increasing collisions energy when using the KMR UGDFs.
Whether this a signal of nonlinear effects (saturation) is in our opinion
still an open issue. We have shown that one can avoid the conflict with 
the LHCb experimental data using UGDFs that include nonlinear effects.

Finally we have addressed shortly the simultaneous production of 
two $J/\psi$ mesons at intermediate transverse momenta. 
We have made calculation relevant for recent ATLAS data. Several 
mechanisms have been discussed. Several distributions have been shown.
The leading (box) contribution was calculated in collinear approach.
The double parton scattering was calculated using a data-driven approach
using a simple parametrization of inclusive $J/\psi$ data.
It seems impossible to describe the new data with the two mechanisms. 
We have also presented results including the two-gluon exchange
mechanism between two quark-antiquark fluctuations of gluons
which turned out to be important at large rapidity difference between
the two $J/\psi$ mesons. Some plans for the future have been presented.



\end{document}